\documentclass[%
reprint,
superscriptaddress,
nofootinbib,
amsmath,amssymb,
aps,
prb,
floatfix,
]{revtex4-2}
\usepackage{amsmath}
\usepackage{amssymb}
\usepackage{mathtools}
\usepackage{adjustbox}
\usepackage{enumitem}
\usepackage{soul}
\usepackage{cancel}
\usepackage{multirow}
\usepackage{tikz}
\usepackage{tabularray}

\usepackage{physics}
\usepackage{bbold}
\usepackage{graphicx}% Include figure files
\usepackage{dcolumn}% Align table columns on decimal point
\usepackage{bm}% bold math
\usepackage[breaklinks]{hyperref}
\hypersetup{colorlinks=true, linkcolor=blue, citecolor=blue, filecolor=blue, urlcolor=blue}

\usepackage[capitalise]{cleveref}

\newcommand{\red}[0]{\color{red}}

\usetikzlibrary{decorations.markings,decorations.text,decorations.pathmorphing,decorations.pathreplacing,
arrows.meta,positioning,intersections,calc,fit}
\usetikzlibrary{mindmap,backgrounds,trees,shapes.geometric}

\NewDocumentCommand{\prong}{m O{solid} O{#2} O{#2}}%
{%
	\foreach \t/\s in {90/#2}
		\draw[midarrow,\s] #1 -- +(\t:1);
	\foreach \t/\s in {210/#3,330/#4}
		\draw[midarrow=reversed,\s] #1 -- +(\t:1.2);
}

\NewDocumentCommand{\pronglabel}{m O{solid} O{#2} O{#2}}%
{%
	\foreach \t/\s in {90/#2}
	\draw[midarrow,\s] #1 -- +(\t:1) node[above] {$c$};
	\foreach \t/\s/\l in {210/#3/a,330/#4/b}
	{	
		\draw[midarrow=reversed,\s] #1 -- node[above=2pt,pos=0.8]{$\l$} +(\t:1.2);
	}
}
\tikzset{->/.style={-{Stealth[]}}}
\tikzset{<-/.style={{Stealth[]}-}}

\tikzset{midarrow/.style={decoration={markings,mark=at position 0.7 with {\arrow{Stealth[#1]}}},
	postaction={decorate}},
midarrow/.default=}

\NewDocumentCommand{\meas}{O{(0,0)}O{1}}{

	\begin{scope}
		\coordinate (o) at #1;
		\draw[fill=white] (o) rectangle node[fill=white] (R) {} +(0.75,0.5);		
		\draw (R) arc [start angle = 90,end angle = 120,radius=0.5];
		\draw (R) arc [start angle = 90,end angle = 60,radius=0.5];
		\draw[-{Stealth[length={#2mm*1.75}]}] ($(o) + (0.375, 0.1)$) -- +(60:0.4);
	\end{scope}
}

\NewDocumentCommand{\repCirc}{O{0.5}O{(0,0)}}{
	\begin{scope}[scale=#1,shift={#2},every node/.style={}]
		\foreach \x in {0,1}
		{
			\draw[xshift=\x cm] (0,-0.25) -- (0,1.5);
		}
		
		\draw[fill=white] (-0.3,0.25) rectangle node {$U$} (1.3,1);
		\meas[(-0.4,1.5)][#1];
	\end{scope}
}

\begin{document}
\title{Entanglement Transitions in Noisy Quantum Circuits on Trees}

        \author{Vikram Ravindranath}
        \email{vikram.ravindranath@bc.edu}
        \affiliation{Department of Physics, Boston College, Chestnut Hill, MA 02467, USA}

    \author{Yiqiu Han}
        \affiliation{Department of Physics and Center for Theory of Quantum Matter, University of Colorado Boulder, Boulder, Colorado 80309, USA}

            \author{Xiao Chen}
	\affiliation{Department of Physics, Boston College, Chestnut Hill, MA 02467, USA}
\begin{abstract}
    Decoherence is ubiquitous, and poses a significant impediment to the observation of quantum phenomena, such as the measurement-induced entanglement phase transition (MIPT). In this work, we study entanglement transitions in quantum circuits on trees, subject to both noise and measurements. We uncover a rich phase diagram that describes the ability of a tree quantum circuit to retain quantum or classical information in the presence of decoherence. By developing a mapping between the dynamics of information on the tree to a classical Markov process -- also defined on the tree -- we obtain exact solutions to the entanglement transitions displayed by the circuit under various noise and measurement strengths. Moreover, we find a host of phenomena, including the MIPT, which are \textit{robust} to decoherence. The analytical tractability facilitated by the method developed in this paper showcases the first example of an exactly solvable noise-robust MIPT, and holds promise for studies on broader, tree-like circuits.
\end{abstract}
\maketitle

\section{Introduction}
Quantum systems appear to undergo two broadly distinct classes of time evolution. The first class is that of unitary dynamics, examples of which include familiar evolution under a Hamiltonian, and quantum gates. The second class involves measurements, which cause abrupt changes to the quantum state. A commonly held belief was that unitary dynamics generically leads to the growth of quantum entanglement between different parts of a system, while measurements -- specifically, projective ones -- would lead to the destruction of entanglement, except in special cases. A new paradigm, however, was ushered in by the discovery of measurement-induced (entanglement) phase transitions, or MIPTs, where unitary dynamics and measurements were found to ``compete" to create or destroy entanglement \cite{MIPT0,MIPT1,MIPT2,MIPT3}.

One of several equivalent interpretations of this phenomenon is the ``purification" picture \cite{Gullans_2020}, which proceeds as follows. Qubits that are initially maximally entangled to some reservoir are subject to a 1+1 dimensional circuit involving both unitary gates, and projective measurements performed at a rate $p$ per qubit per time step. It was shown that below a critical rate $p_c$, the system remains entangled to the reservoir, while above this rate, the system becomes completely disentangled. Further interpreting the measurements as ``errors" led to a physical understanding of the MIPT as a dynamically generated quantum error correcting code, in which the unitary gates scramble the quantum information, thereby ``protecting" from leakage due to local measurements \cite{Gullans_2020,Choi_2020,Fan_2021,Li_2021}. Such transitions have most commonly been quantified by the scaling of the entanglement entropy of the system, where it scales with the volume (area) of the system below (above) $p_c$. On the theoretical front, the MIPT has been explored via a mapping to classical spin chains, defined in replica space (involving multiple copies of a system) \cite{MIPTheo1,MIPTheo2}. The entanglement entropy of a subsystem corresponds here to the free energy cost of introducing a domain wall on either end of the subsystem at one boundary. In 1+1 D, the MIPT can then be mapped to a 2+0 D ferromagnetic-paramagnetic phase transition, where $p$ plays the role of temperature; in the ferromagnetic phase, the free energy cost is linear in the size of the subsystem considered.

In addition to unitary dynamics and projective measurements, interactions with the environment present a third, qualitatively different class of dynamics. These interactions can (and often do) lead to noisy decoherence, as they generically cause the evolution of the system into a partially or completely unknown state. Since these are ubiquitous in nature, it is of interest to understand how the MIPT is affected, if at all, by noise. Unfortunately, this has been answered in the negative in 1+1 dimensional circuits; the addition of vanishingly small noise tends to wash out the MIPT, leaving the system unable to protect any information \cite{Noise1,Noise2,Noise3,Li_Hsieh}. One physical explanation for this arises from the classical mapping -- noise (specifically, the action of a depolarizing channel) plays the role of a symmetry-breaking magnetic field. It is well known that ferromagnetic-paramagnetic transitions are absent in the presence of arbitrarily small symmetry breaking field. One might naturally wonder if this is generically the case.

In this paper, we explore this question and find a setting in which the MIPT persists in the presence of noise. We discover a rich phase diagram in which transitions between different entanglement phases — representing different capacities to preserve quantum or classical information — are not only robust to noise but can even be viewed as \textit{driven} by noise. We consider a family of quantum circuits defined on the tree. Our motivation draws from the physics of magnetic transitions on trees, as opposed to square lattices or other high-dimensional regular lattices. The Ising spin model on a tree can exhibit a finite-temperature phase transition if the boundary spins are constrained, and this transition can persist even in the presence of an external magnetic field. This robustness serves as inspiration for exploring the existence of MIPTs under conditions of decoherence.

Building on this picture, we examine hybrid Clifford circuits on a tree structure subject to depolarization noise. To characterize different quantum phases, we utilize the mutual information between the leaves and the root of the tree. Using the unique properties of Clifford circuits and the recursive structure of trees, we demonstrate that this problem can be mapped onto an effective classical Markov process. By analyzing the associated master equation, we identify distinct phases and determine thresholds for both noise and measurement rates.

Before proceeding, we note that recent works have taken different approaches to addressing this problem \cite{qian2024coherent,qian2024protect,kelly2024,kelly2024generalizing}, by using more nuanced quantum information measures. In parallel, other constructions on trees \cite{akshar2023noisy,sommers2024dynamically} have been studied, which examine the ability of non-random circuits to protect information, by explicitly constructing decoding algorithms. Lastly, we note that an analytical solution for the MIPT on a tree was previously obtained \cite{feng2023tree} in the absence of noise. There, the authors considered a circuit consisting of weak measurements and unitary gates drawn from the Haar distribution, and obtained an analytical solution, albeit one that required linearization. In our work, by considering Clifford dynamics and a mapping to a Markov process, we are able to obtain completely exact results for the MIPT on the tree, thereby determining the critical exponents and the measurement rates exactly.

The rest of this paper is organized thus. Our setup is detailed in \cref{sec:setup}, followed by the motivations behind our considerations of trees. This is done by reviewing results regarding the Ising model on the tree and a simple example from classical error correction. \cref{sec:mipt} deals with exact solutions and results for the MIPT in the absence of noise, while the introduction of noise and a complete phase diagram are discussed in \cref{sec:noise}. Finally, we discuss the extension of our results to more involved setups in \cref{sec:ext}, and present a summary of our results in \cref{sec:conc}.

\section{Setup and Motivation}
\label{sec:setup}
Our setup consists of tree quantum circuits of variable depth $D$, constructed using the following prescription. We begin by considering $k^D$ qubits, each of which forms a Bell pair with another ancilliary qubit, for a total of $2k^D$ qubits. These ancilliary qubits will henceforth be referred to as ``leaves" and denoted $L$, and do not participate further in the quantum circuit. The remaining qubits are collected into groups of $k$ qubits each, and each group is subject to a $k-$qubit Clifford gate. Following this, $k-1$ qubits in each group are measured in the $Z$ basis and discarded, leaving $k^{D-1}$ qubits.  The resulting setup emulates the purification dynamics of a maximally entangled state subject to a tree quantum circuit that involves measurements and unitary gates \cite{Gullans_2020}. 

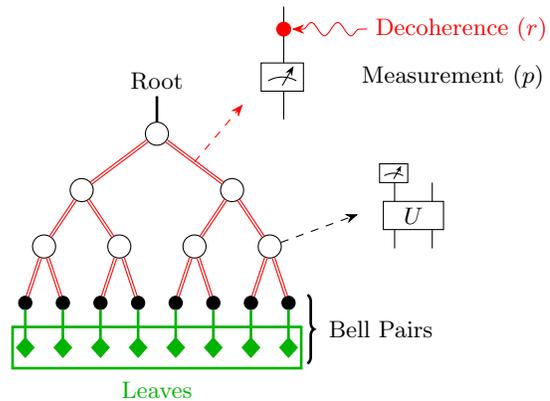
\begin{figure}
    \centering
    \begin{tikzpicture}[level distance=0.75cm]
        
        \tikzset{edge from parent/.append style={red,double},
            %edge from parent path={(\tikzparentnode\tikzparentanchor) [red,double,fill=red] -- (\tikzchildnode\tikzchildanchor)},
            every node/.style={draw, color=black,circle,fill=white,minimum width=5pt},
            level 1/.style={sibling distance=2cm},
            level 2/.style={sibling distance=1cm},
            level 3/.style={sibling distance=0.5cm, nodes={draw,fill,color=black,inner sep=1pt}}}
        \draw[line width=1pt] (0,0) -- (0,0.5) node[draw=none,above,inner sep=-5pt] {Root};
        \node [circle,draw,fill=white] (preR) at (0,0) {}
        child {node {} 
            child { node {}
                child {node (a1) {}}
                child {node (a2) {}}
            }
            child { node  {} 
                child {node (b1) {}}
                child {node (b2) {}}
            }
        }
        child {node (cd) {}
            child { node (c) {} 
                child {node (c1) {}}
                child {node (c2) {}}
            }
            child { node (d){} 
                child {node (d1) {}}
                child {node (d2) {}}
            }
        };
        
        \foreach \x in {a,b,c,d}
        {
            \foreach \y in {1,2}
            {
                \draw[color=green!70!black,line width=1pt] (\x\y.south) -- +(0,-0.5cm) node[draw=none,diamond, aspect=0.9,fill=green!70!black,inner sep=2pt] (b\x\y) {};
            }
        };
        
        \draw[decorate,decoration={brace},line width=1pt] let \p{d2e}= (d2.east),\p{bd2e} = ($(bd2.east) +(0.1,0)$) in
        (\x{bd2e},{\y{d2e}+3}) -- (\x{bd2e},{\y{bd2e}-6}) node[draw=none,fill=none,midway,xshift=3em] {Bell Pairs};

        \begin{scope}[label distance=-10pt]
            \node[rectangle,fit = (ba1) (bd2),draw,line width=1,color=green!70!black,fill=none,label=below:{\color{green!70!black}Leaves},inner xsep=0.5] {};
        \end{scope}
        
        \draw[dashed,->] (d) -- +(20:1.25cm);
        
        \repCirc[0.5][($(d.east)+(3,0.2)$)];

        \path[blue,->,name path=p1] (preR) -- (cd);
        \path[blue,->,name path=p2] ($0.5*(preR) + 0.5*(cd)$) -- +(0.2,0);
        \draw[red,->,dashed,name intersections={of=p1 and p2}] (intersection-1) -- +(50:1cm);
        \coordinate (wire) at ($(intersection-1) + (50:0.25cm) + (-0.1,0.75)$);

        \begin{scope}[scale=0.75,shift=(wire),xshift=1.5cm,yshift=-0.5cm,pin distance = 1cm,
            every pin edge/.style={<-,decorate,
                decoration={snake,pre length=1.5mm},color=red}]
            \tikzset{every node/.style={fill=white}}
            \draw (0,0) -- (0,2);
            \meas[(-0.4,0.5)][0.75];
            \node[draw=none,inner sep=2pt] at (3cm,0.75) {Measurement $(p)$};
            \path (0,0) -- node[circle,fill=red,pos=0.8,inner sep=2pt,pin=0:{\color{red}Decoherence} $\color{red}(r)$]{} (0,2);
            
        \end{scope}
        
    \end{tikzpicture}
    
    \caption{A schematic of the circuit set-up, illustrated for a depth of $D=3$ with $k=2$. $k^D$ qubits are initialized in Bell pairs with an equal number of qubits called the leaves, and the circuit does not act further on these leaves. Each  qubit passes through a noisy wire (denoted by \protect\tikz[baseline]{\protect\draw[red,double] (0,2.5pt) -- (0.25,2.5pt) ;}) where it independently undergoes projective measurements with probability $p$ and decoherence at rate $r$. Qubits are then collected in groups (or ``nodes", denoted by white unfilled circles) of $k=2$, to which $k-$qubit Clifford gates are applied, and all but one qubit are measured and traced out. This process subsequently repeats until only a single qubit -- the root -- remains.}
    \label{fig:schem1}
\end{figure}

In each subsequent layer, the remaining qubits are collected into groups of $k$, and this procedure is iterated until only 1 qubit remains -- the ``root" $R$. Between layers, each qubit is subject to noise at a rate $r$ and projective measurements at a rate $p$. For much of this work, we will restrict our attention to $k=2$, as illustrated schematically in \cref{fig:schem1}.

The noise model that we consider is a specific unraveling of the depolarizing channel acting on individual qubits. Its action on a specific qubit $j$ is given by
\begin{equation*}
    \rho \to (1-r)\rho + r \frac{I_j}{2}\otimes\Tr_j(\rho).
\end{equation*}
This is implemented by tracing each qubit out and replacing it by a maximally mixed state, independently, at a rate $r$. This unraveling also has an equivalent interpretation of maximally coupling each tree qubit to an environment qubit (with rate $r$), following which the environment qubit is discarded (traced out).

Explicitly, let two qubits labelled $a$ and $b$ meet at a node, which is a combination of a $2$-qubit gate and a measurement, shown as white unfilled circles $\bigcirc$ in \cref{fig:schem1}. Their joint state is $\rho_{a,b,L}$, since $a$ or $b$ could be correlated with the leaves $L$. At the node,

\begin{equation}
    \rho_{a,b,L} \to \rho'_{b,L} \propto \Tr_a\qty[\mathbb{P}_a U_{ab} \rho_{a,b,L} U^\dagger_{ab} \mathbb{P}_a].
    \label{eq:dyn_rule}
\end{equation}
$U_{ab}$ is a unitary gate that acts on $a$ and $b$. $\mathbb{P}_a \equiv \frac{1\pm Z}{2}$ describes a projective measurement on qubit $a$, following which it is traced out. In the following, we also occasionally refer to the output state as $\rho_{c,L}$, i.e.\ we relabel the qubit at the output as $c$ to underline the fact that the output qubit could physically either be $a$ or $b$, but which specific qubit is eventually irrelevant.

In this work, we are primarily interested in determining if information encoded between the leaves and the root, characterized by the mutual information, survives in the limit of $D\to\infty$ in the presence of environmental noise and measurement. The mutual information is defined as
\begin{align}
    I(R;L) = S_R + S_L - S_{RL}
\end{align}
where
\begin{align*}
    S_A = -\Tr{\rho_A \ln (\rho_A)},
\end{align*}
and
\begin{align*}
    \rho_A \equiv \Tr_{\overline{A}}(\rho)
\end{align*}
is the reduced density matrix on system $A$, obtained by tracing out $\overline{A}$, the complement of $A$, from the density matrix $\rho$ describing the complete system. Unlike standard MIPT setups, where environmental noise is absent, entanglement entropy alone is no longer a suitable order parameter for detecting entanglement transitions. Decoherence at any nonzero rate contributes to the entropy of the root qubit, causing $S_A$ to remain smooth and finite as a function of $r$. Thus, in order to detect genuine correlations (classical or quantum), we must calculate $I$ instead, which is specifically defined to exclude such spurious contributions.

\subsection{Transitions in the Tree Ising Model}
\label{ssec:ising}
As noted in the introduction, the physics underlying entanglement transitions can be heuristically captured by a mapping between the quantum circuit and a classical Ising spin model defined on a tree. The punchline here is that the free energy cost of introducing a domain wall between the roots and the leaves undergoes a transition with temperature even in the presence of a magnetic field. This leads us to conjecture (and subsequently show) that quantum circuits on trees should show an MIPT even in the presence of noise.

Building on established results regarding the mapping between an MIPT and the order-to-disorder thermal phase transition in a classical interacting spin model~\cite{MIPTheo1,MIPTheo2}, the unitary gates result in pairwise couplings between every pair of vertices joined by an edge.\footnote{Strictly speaking, such a mapping was only rigorously shown for gates drawn from a Haar ensemble. However, since the Clifford group forms a 3-design \cite{webb20163design}, the calculation of the second R\'enyi entropy $S^{(2)}_A\equiv -\ln\qty(\Tr \rho_A^2)$ is insensitive to which of the two ensembles was used. Clifford states are also special in that their R\'enyi entropies are identically equal to their von Neumann entropy $S_A$. This allows us to draw qualitative conclusions about the physics of random Clifford circuits, from the same classical considerations.} The measurement rate $p$ plays the role of temperature $T$, and $r$ is related to the magnetic field $h$. It is known that the Ising model on a tree is somewhat special and exhibits ferromagnetic-paramagnetic transitions only when the boundary spins are constrained. However, this poses no impediment, since such boundary conditions are naturally imposed in the calculation of the mutual information $I$.

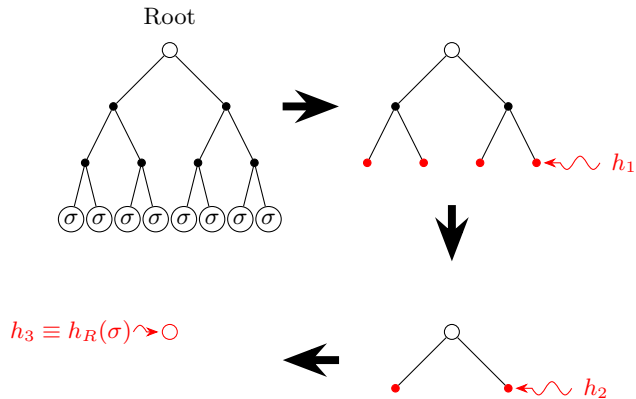
\begin{figure}
    \centering
    \begin{tikzpicture}[scale=0.75,
    every node/.style={circle, draw,fill=black,inner sep=1pt},
			level distance=1cm,
			level/.style={sibling distance=4cm/(2^(#1))},
			level 3/.style={every node/.style={circle, draw, inner sep=1pt, node contents={$\sigma$}}}]
			\node[circle,fill=none, inner sep=2pt, label={[label distance=-2pt]above:Root}] at (0,0) (R0) {}
			child { node {}
				child { node {}
					child {node {}}
					child {node {}}
				}
				child { node {}
					child {node {}}
					child {node {}}
				}
			}
			child { node {}
				child { node {}
					child {node {}}
					child {node {}}
				}
				child { node {}
					child {node {}}
					child {node {}}
				}
			};
						
			\begin{scope}[xshift=5cm,
				level 2/.style={every node/.append style={fill=red,draw=red}}]
				\node[circle,fill=none, inner sep=2pt, label={[label distance=-2pt]above:}] (R1) at (0,0) {}
				child { node {}
					child {node {}}
					child {node {}}
				}
				child { node {}
					child {node {}}
					child {node {}}
				};
				\draw[->={1mm},decorate,decoration={coil,aspect=0},red] ($(R1-2-2.west) + (1.2,0)$) node[draw=none,fill=none,label=0:$h_1$]{} -- ($(R1-2-2.west) + (0.2,0)$);
			\end{scope}
			
			\begin{scope}[xshift=5cm, yshift=-5cm,
				level 1/.style={every node/.append style={fill=red,draw=red}}]
				\node[circle,fill=none, inner sep=2pt, label={[label distance=-2pt]above:}] (R2) at (0,0) {}
				child {node {}}
				child {node {}};
				\draw[->={1mm},decorate,decoration={coil,aspect=0},red] ($(R2-2.west) + (1.2,0)$) node[draw=none,fill=none,label=0:$h_2$]{} -- ($(R2-2.west) + (0.2,0)$);
			\end{scope}
			
			\begin{scope}[yshift=-5cm,
				level 1/.style={every node/.append style={fill=red,draw=red}}]
				\node[circle,fill=none, draw=red, inner sep=2pt, label={[label distance=-2pt]above:}] (R3) at (0,0) {};
				\node[fill=none,draw=none,inner sep=0,color=red] at (-1.75,0) (h) {$h_3\equiv h_R\qty(\sigma)$};
				\draw[->={1mm},decorate,decoration={coil,aspect=0},xshift=-0.2cm,red] (h.east) -- (0,0);
			\end{scope}

			\draw[->={5mm},line width=3pt] ($(R0-2)+(1,0)$) -- ($(R1-1)-(1,0)$);
			\draw[->={5mm},line width=3pt] ($($(R1-1-2)! 0.5!(R1-2-1)$) + (0,-0.75)$) -- +(0,-1);
			\draw[<-={5mm},line width=3pt] let \n1={-4.5} in ($(R0-2)+(1,\n1)$) -- ($(R1-1)+(-1,\n1)$);
			
	\end{tikzpicture}
    \caption{A diagramatic representation of how spin models on trees are solved is shown for a tree of depth $D=3$. The spins on the leaves are fixed to be in the configuration $\sigma$. Spins connected by an edge interact through an Ising coupling. Nodes in red are subject to a magnetic field induced by spins located one level deeper; this inducement of a field results from performing the summation over the spins at the greatest depth at each iteration, with the specific values of $h_D$ given by \cref{eq:rec_ising}. Ultimately, the model reduces to the calculation of the magnetic field on the root, which we term $h_R(\sigma)$ for a given leaf configuration $\sigma$.}
    \label{fig:tree_solve_schem}
\end{figure}

The calculation of the entanglement entropy $S_A$ of a region $A$ on the periphery of the tree (corresponding to the time when the circuit has terminated) reduces to the calculation of free energies of the spin model on the tree under different boundary conditions for the spins on the periphery
\begin{equation}
    S_A = F\qty(\downarrow_A\uparrow_{\overline{A}})-F\qty(\uparrow_A\uparrow_{\overline{A}})
    \label{eq:S_MF}
\end{equation}
where $\sigma_A$ denotes the configuration where all the spins on region $A$ are constrained to be $\sigma \in \qty{\uparrow,\downarrow}$. On the tree, when $A=R$ (and $\overline{A} = L$), the free energies \cref{eq:S_MF} have a particularly simple interpretation: $F(\downarrow_R\sigma_L) - F(\uparrow_R\sigma_L) \equiv h_R(\sigma_L)$ is the strength of the effective magnetic field (owing to all the other spins) on the root, when the spins of leaves are fixed in the configuration $\sigma_L\in\qty{\uparrow,\downarrow}$. Note that the analogous constraint on $\sigma_R$ is no longer present, since the two configurations of $\sigma_R$ have already been summed over in the definition of $h_R(\sigma_L)$. The mutual information reduces to 

\begin{equation}
    \begin{aligned}
        I(R;L) &= S_R + S_L - S_{RL}\\
        &= F(\downarrow_R\uparrow_L) - F(\uparrow_R\uparrow_L)\\
        &+ F(\uparrow_R\downarrow_L) - \cancel{F(\uparrow_R\uparrow_L)}
        -F(\downarrow_R\downarrow_L) + \cancel{F(\uparrow_R\uparrow_L)}\\
        I &= h_R(\uparrow) - h_R(\downarrow)\equiv \Delta h_R.
    \end{aligned}
    \label{eq:def_DH}
\end{equation}
The mutual information $I$ corresponds to the difference $\Delta h_R$ in the magnetic field strength at the root, when the leaf spins are in opposite orientations. We now calculate $I$ for the Ising model on the tree, at inverse temperature $\beta$, subject to a uniform magnetic field $h\geq0$. $I$ can be explicitly calculated by mapping the partition function of a tree of depth $D$\footnote{As measured from the base/leaves of the tree} to one of depth $D-1$, albeit with magnetic fields on the leaves $h_{D-1}$ to account for the interactions arising from interactions with the spins of the tree of depth $D$. This process is depicted in \cref{fig:tree_solve_schem}. For completeness, we present this process in more detail in \cref{app:treeising}.

\begin{figure}
    \centering
    	\begin{tikzpicture}[scale=0.75,
        every node/.style={circle, draw,fill=black,inner sep=1pt},
		level distance=1cm,
		level/.style={sibling distance=4cm/(2^(#1))},
		level 3/.style={every node/.style={circle, draw, inner sep=1pt, node contents={\tikz\draw[->={3pt}](0,0) -- (0,5pt);}}}]
		\node[circle,fill=none, inner sep=2pt] at (0,0) {}
		child { node {}
			child { node {}
				child {node {}}
				child {node {}}
			}
			child { node {}
				child {node {}}
				child {node {}}
			}
		}
		child { node {}
			child { node {}
				child {node {}}
				child {node {}}
			}
			child { node {}
				child {node {}}
				child {node {}}
			}
		};
		
		\node[fill=none,draw=none,inner sep=0] at (-1.75,0) (h) {$h_R\qty(\uparrow)$};
		 \draw[->={1mm},decorate,decoration={coil,aspect=0},xshift=-0.2cm] (h.east) -- (0,0);

		 \begin{scope}[xshift=5cm,
		 	level 3/.style={every node/.style={circle, draw, inner sep=1pt, node contents={\tikz\draw[->={3pt}](0,0) -- (0,-5pt);}}}]
		 	\node[circle,fill=none, inner sep=2pt] at (0,0) {}
		 	child { node {}
		 		child { node {}
		 			child {node {}}
		 			child {node {}}
		 		}
		 		child { node {}
		 			child {node {}}
		 			child {node {}}
		 		}
		 	}
		 	child { node {}
		 		child { node {}
		 			child {node {}}
		 			child {node {}}
		 		}
		 		child { node {}
		 			child {node {}}
		 			child {node {}}
		 		}
		 	};
		 	
		 	\node[fill=none,draw=none,inner sep=0] at (-1.75,0) (h) {$h_R\qty(\downarrow)$};
		 	\draw[->={1mm},decorate,decoration={coil,aspect=0},xshift=-0.2cm] (h.east) -- (0,0);
		 \end{scope}
	\end{tikzpicture}
    \[I(R;L)\sim \Delta h_R = h_R(\uparrow) - h_R(\downarrow)\]
    \caption{The calculation of $I$ is qualitatively captured by the difference in the magnetic field at the root $\Delta h_R$, when the leaves are constrained in opposite configurations.}
    \label{fig:Ising_MI_schem}
\end{figure}
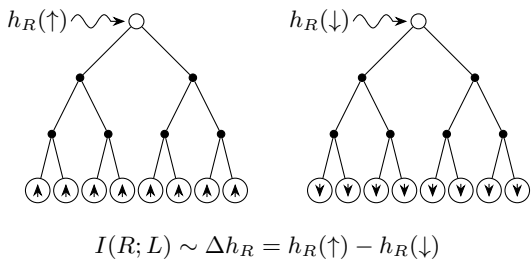

\begin{figure*}
    \centering
    \includegraphics[width=0.9\textwidth]{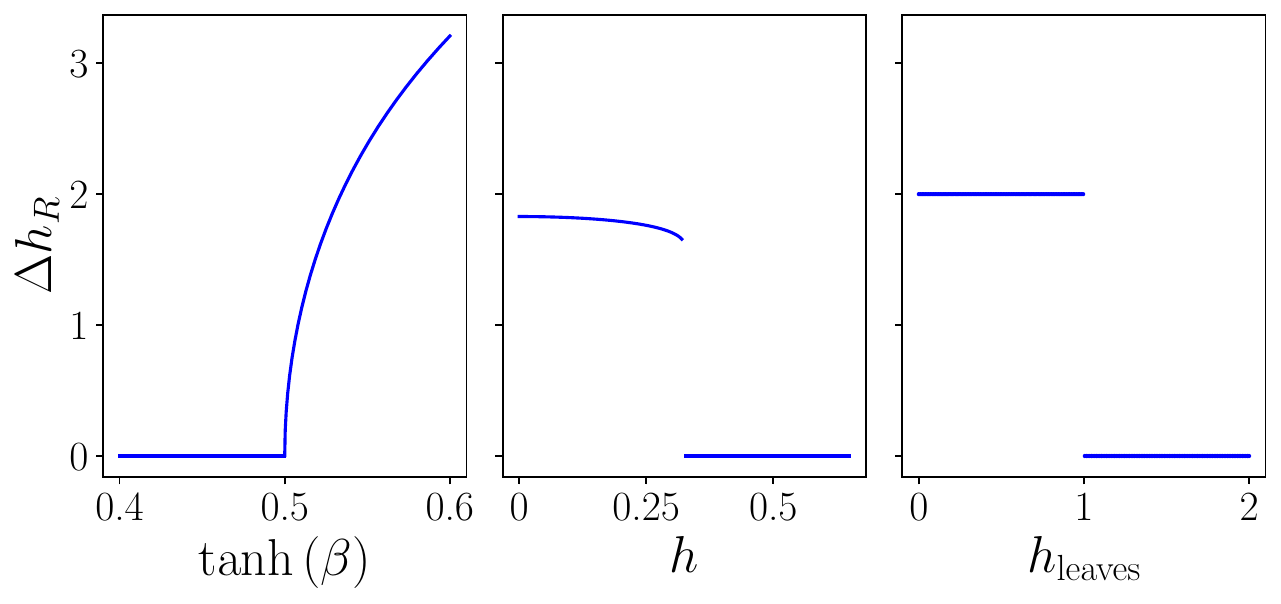}
    \caption{The difference in the magnetic field at the root $\Delta h_R$, plotted as a function of the inverse temperature (left), the global magnetic field (center) and the boundary magnetic field (right), in the limit of infinite depth $D\to\infty$. (Left) When $h=0$, there is a continuous phase transition at $\tanh(\beta)=\frac{1}{2}$ between a low temperature phase where $h_R\neq0$ to one where $h_R=0$. (Center) At $\beta=1$, $h\neq0$, there is now a \textit{first-order} as the magnetic field $h$ is tuned through $h_c\approx0.323$. (Right) Lastly, when $\beta$ and $h$ are tuned to ensure that the system is in the $\Delta h_R\neq0$ phase, the magnetic fields on the leaves can be tuned to independently drive a transition in the root magnetic field. Here, $\beta=10$ and $h=0.3$. Increasing $h$ drives the location of $h^c_{\rm leaves}\to0$.}
    \label{fig:IsingMI}
\end{figure*}

Firstly, when $h=0, h_R(\uparrow) = -h_R(\downarrow)$, owing to the $\mathbb{Z}_2$ Ising symmetry, so $I=2h_R(\uparrow) = 2S_R$. When the leaf spins are constrained to all be $\sigma$, the Ising model on the tree undergoes a second-order thermal phase transition, reflected in the root magnetization. For $T<T_c, h_R \neq 0$, whereas $h_R\to0$ when $T>T_c$. As we will show, the entanglement entropy rightly undergoes a transition as $p$ is varied. The transition in $\Delta h_R$ is shown in the left panel of \cref{fig:IsingMI} as a function of the inverse temperature $\beta\equiv1/T$.

When $h>0$, such a transition still persists, but is now a qualitatively different first-order phase transition. For small $h$, the effect of the leaf spins dominates the effect of the field, leading to a non-zero contribution to $I$. However, above a value $h_c$, the leaf spins' magnetic field no longer influences the root spin, and $h_R(\uparrow) = h_R(\downarrow) \sim h$, and thus, $I=0$. This is shown in \cref{fig:IsingMI} (center), where $\beta$ is fixed and the global magnetic field $h$ is varied. The transition is a first-order transition even if $0<h<h_c$ and $\beta$ is tuned instead.

Lastly, a special feature of the Ising model on trees is that transitions can arise from tuning the boundary fields (i.e.\ the fields on the leaves) alone. In the presence of a global magnetic field at finite temperature, the root magnetization behaves non-trivially as the boundary magnetic field is varied. In the analogy to quantum circuits, a boundary magnetic field corresponds to noise purely on the leaf qubits. The results of \cref{fig:IsingMI} provide an insight into the phase diagram of a family of random quantum circuits on the tree, when the qubits are subject to noise and projective measurements, even if noise is asymmetrically applied globally and on the leaves.

Armed with this intuitive backing, the final ingredient that will enable the calculation of $I$ is a classical Markov process, also defined on the tree.

\subsection{Effective Classical Markov Process}

A property of Clifford gates that we will use extensively is that the entanglement of Clifford states can only be nonnegative integral multiples of $\log 2$ -- it is discrete. Moreover, the root of the tree, by virtue of being a single qubit, can exist in a joint state with the leaves have 3 values of $I$ -- $I(R;L) = 2, 1$ or $0$. When $I=0$, the root can be a mixed $\rho_R \propto \mathbb{1}$ (denoted $M$) or a pure state $\rho_R = \frac{1+\sigma}{2}$ where $\sigma = X, Y$ or $Z$ Pauli matrices. With this in mind, we consider a 4 dimensional ``c-state" space (where ``c" stands for ``classical") on which a dynamical process can be defined. 2 of these c-states will be labelled $2$ and $1$, corresponding to the cases where the root has a mutual information $I(R;L)$ with the leaves of $2$ or $1$ respectively. The remaining $I=0$ states will be labelled $M$ and $\sigma$\footnote{The unitary ensembles under consideration for the most part are invariant under single qubit rotation, so pure states will be identically labelled as $\sigma$, regardless of whether their stabilizers are $X, Y$ or $Z$.}. Thus, after averaging over locations where decoherences and measurements can act, over gate choices and over possible inputs, the root at a specific depth $D$ will be characterized by a probability distribution $P_D(c)$, where $c \in \qty{2, 1, M,\sigma}$ is its c-state. The process is Markovian since the probability $P_{D+1}(c)$ of obtaining a state $c$ at depth $D+1$ depends only on the probability distribution  $\qty{P_D(c')}$ at depth $D$.

Beginning from the base of the tree, the inputs to each node at height $D$ can be interpreted as the roots of \textbf{independent} trees of depth $D-1$. The root (or output) at height $D$ will undergo a measurement with probability $p$, and be subject to decoherence with probability $r$. Subsequently, it will serve as an input to a node at height $D+1$ alongside $k-1=1$ other such root(s), and the process is repeated. This recursive tree structure enables the derivation of a recursion relation, as we now show. By determining the transition matrix $W((a,b)\to c)$ for c-states $a$ and $b$ to produce $c$, we have the recursion relation

\begin{equation}
    P_{D+1}(c) = \sum\limits_{\qty{a_j}_{j=1}^k} W((\qty{a_j})\to c) \qty(\prod_j P_D(a_j)).
    \label{eq:gen_rec}
\end{equation}

The steady-state behavior can be obtained from the fixed point probabilities $P_{D\to\infty}(c)\equiv P(c)$. The final, ensemble-averaged mutual information $I$ is then given by

\begin{equation}
    \overline{I_\infty(R;L)} = 2 P_\infty(2) + P_\infty(1).
\end{equation}

A subtlety that needs addressing is that the number of leaves also grow with the depth of the circuit. However, $I(R;L)\leq2$, regardless of the size of $L$, since the mutual information $I(A;B)$ between two subsystems $A$ and $B$ is bounded by the minimum of the Hilbert space dimensions of $A$ and $B$. Therefore, one can identify a two dimensional subspace (or equivalently, the Hilbert space of a single ``logical" qubit), labelled $\mathcal{H}_{2}$, in the Hilbert space of $L$ that captures all the entanglement between $R$ and $L$. Since we only calculate the entanglement between the root, and the leaves as a whole, the leaves for each root $R$ can be replaced by a ``logical" qubit $L'$ whose Hilbert space is $\mathcal{H}_2$. For instance, $I(R;L)=2$ for a given root means that a Bell pair exists between $R$ and $L'$

\begin{equation}
    \ket{\psi}_{R,L} = \frac{1}{\sqrt{2}}\Big(\ket{0}_R\ket{0}_{L'} + \ket{1}_R\ket{1}_{L'}\Big).
    \label{eq:eg_I2}
\end{equation}
On the other hand, if $I(R;L)=1$, this means that the root, leaves and the environment (denoted by $E$) form an effective GHZ state
\begin{equation}
    \ket{\psi}_{R,L,E} = \frac{1}{\sqrt{2}}\Big(\ket{0}_R\ket{0}_{L'}\ket{0}_{E} + \ket{1}_R\ket{1}_{L'}\ket{1}_{E}\Big).
    \label{eq:eg_I1}
\end{equation}

\begin{figure*}[t]
    	\begin{tikzpicture}[level distance=0.75cm,
		every node/.style={draw, solid, color=black,circle,fill=white,minimum width=5pt},
		level 1/.style={sibling distance=2cm},
		level 2/.style={sibling distance=1cm},
		level 3/.style={sibling distance=0.5cm, nodes={draw,solid,fill,color=black,inner sep=1pt}}]
		
		\draw[line width=1pt] (0,0) -- (0,0.5) node[draw=none,above,inner sep=-5pt] {Root};
		\node [circle,draw,fill=white] (preR) at (0,0) {}
		child {node {} 
			child { node {}
				child {node (a1) {}}
				child {node (a2) {}}
			}
			child { node  {} 
				child {node (b1) {}}
				child {node (b2) {}}
			}
		}
		child {node (cd) {}
			child { node (c) {} 
				child {node (c1) {}}
				child {node (c2) {}}
			}
			child { node (d){} 
				child {node (d1) {}}
				child {node (d2) {}}
			}
		};
		
		\foreach \x in {a,b,c,d}
		{
			\foreach \y in {1,2}
			{
				\draw[color=green!70!black,line width=1pt] (\x\y.south) -- +(0,-0.5cm) node[draw=none,diamond, aspect=0.9,fill=green!70!black,inner sep=2pt] (b\x\y) {};
			}
		};

		\begin{scope}[xshift=5cm]
			\draw[line width=1pt] (0,0) -- (0,0.5) node[draw=none,above,inner sep=-5pt] {Root};
			\node [circle,draw,fill=white,label={[label distance=-2pt]176:$n_0$}] (preR) at (0,0) {}
			child {node[label={[label distance=-4pt]135:$n_1$}] {} edge from parent [solid,red]
				child { node [label={[label distance=-4pt]135:$n_2$}] {} edge from parent [solid,black]
					child {node (a1) {} edge from parent [red,dashed]}
					child {node [label={[label distance=-3pt]45:$n_3$}](a2) {}}
				}
				child { node  {} edge from parent [red,dashed]
					child {node (b1) {} edge from parent [solid,black]}
					child {node (b2) {} edge from parent [solid,black]}
				}
			}
			child {node (cd) {} edge from parent[red,dashed]
				child { node (c) {} edge from parent [red,solid]
					child {node (c1) {} edge from parent [dashed]}
					child {node (c2) {} edge from parent [solid,black]}
				}
				child { node (d){} edge from parent[red,dashed] 
					child {node (d1) {} edge from parent [red,dashed]}
					child {node (d2) {} edge from parent [red,dashed]}
				}
			};
			
			\foreach \x in {a,b,c,d}
			{
				\foreach \y in {1,2}
				{
					\draw[color=green!70!black,line width=1pt] (\x\y.south) -- +(0,-0.5cm) node[draw=none,diamond, aspect=0.9,fill=green!70!black,inner sep=2pt] (b\x\y) {};
				}
			};
		\end{scope}
		
		\begin{scope}[xshift=9cm,line width=1pt,yshift=5pt,
			every node/.style={fill=none}]
			\path (0,0.5) -- +(0.75,0) node [right=0.2] {\underline{c-states}};
			\draw (0,0) -- (0.75,0) node [right=0.5] {$2$};
			\draw[red] (0,-1) -- (0.75,-1) node [color=black,right=0.5] {$1$};
			\draw[dashed] (0,-2) -- +(0.75,0) node[right=0.5]{$\sigma$};
			\draw[red,dashed] (0,-3) -- +(0.75,0) node[color=black,right=0.5] (M) {$M$};			
		\end{scope}		
	\end{tikzpicture}
    \caption{A demonstration of the mapping between the circuits that we consider, and a generalized directed percolation of particles from the leaves to the root of the tree, for trees of depth $D=3$. (Left) In the absence of measurements and noise ($p=r=0$), each leaf has an unbroken path (denoted by $I=2$) to the root, so that quantum information can be recovered at the root. (Right) When the circuit is subject to noise, certain bonds are broken (i.e.\ in the state $M$), resulting in ``imperfect" bonds (corresponding to $I=1$). In this particular realization, there is an imperfect path from the root to the leaves. While the bonds from $n_3\to n_1$ are unbroken, allowing quantum information to travel upto $n_1$, the subsequent bond from $n_1\to n_0$ is imperfect, thereby only allowing classical information to travel to the root.}
    \label{fig:GenDP}
\end{figure*}
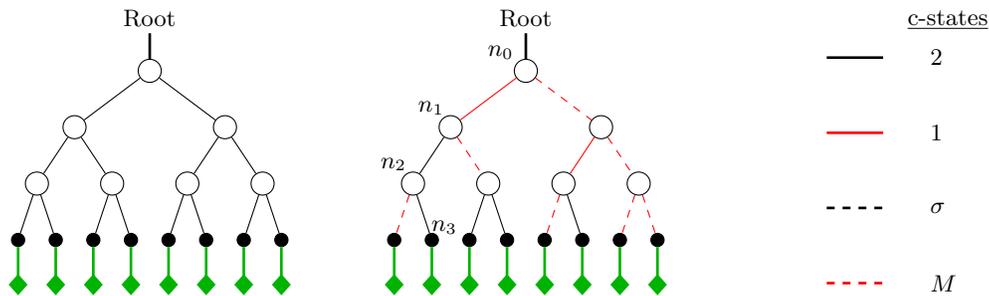

In our results, we present not the mutual information but the probabilities of various values of $I$ themselves. In such Clifford states, $P(I = 2 (1))$ corresponds to the probability that quantum (only classical) information is stored in trees of large depth, and therefore offer more nuanced results from which $I$ can be calculated. It is also conceivable that classical information might have a greater tolerance to noise. We say that quantum/classical information is ``retained" if there is a finite probability, in the limit $D\to\infty$, that $P(I=2/1)\neq0$; alternatively, no information is retained if $P(I=0)\equiv P(M) + P(\sigma)=1$.

Therefore, we have reduced our problem to the calculation of the transition matrix $W$ which can either be done analytically or numerically, and subsequently, solving \cref{eq:gen_rec}, which can be done efficiently. Boundary perturbations are incorporated into the initial conditions $P_{D=0}(c)$, as in \cref{eq:rec_ising}. For instance, if a fraction $f$ of the initial states are pure states, while the remaining are maximally entangled to the leaves, we would have $P_0(\sigma) = 1 - P_0(2) = f$.

A graphically appealing interpretation of this circuit arises from a mapping to a generalized percolation process. We note that, through various mappings to percolation \cite{sang2021mphase,iaco2020autom,skinner2020maj,MIPT2}, the quantification of entanglement between two regions in a system subject to random unitary gates and projective measurements has been reduced to determining an unbroken path between them. In particular, the ``min-cut" method pioneered in \cite{MIPT2} to calculate the Hartley entropy of a system subject to a random quantum circuit with measurements treated the unitary gates as ``bonds", while the measurements effectively ``broke" these bonds. The entanglement entropy of a subsystem was then shown to be proportional to the length of the path that cut through the least number of bonds, and the MIPT was found to obey physics similar to, but not exactly, that of percolation.

A more concrete connection to \textit{directed} percolation (DP) was found in the calculation of the second R\'enyi entanglement entropy in the context of random quantum automaton circuits \cite{iaco2020autom}. For an initial 1D state consisting of qubits maximally entangled to an equal number of ancillas, the mapping proceeds as follows: starting from a 2D lattice completely filled with particles at the top, and empty everywhere else, the particles are allowed to ``percolate" downwards -- this is the effect of the unitary gates. On the other hand, measurements block accessible paths in this lattice. When the measurement rate $p<p_c$, there exist paths to reach the other end, while for larger $p$, all possible paths are broken, thereby causing the particles to be arrested at an $O(1)$ depth into the lattice. The mapping that we describe below allows us to understand entanglement transitions on the tree in a similar vein.

\subsubsection{Generalized Directed Percolation Mapping}

A paradigmatic example of DP on the tree is that of bond directed percolation, the setup for which follows straightforward update rules. Particles are allowed to only travel ``up" the tree, that is, from the leaves towards the root, along the edges of this tree. At each node, an edge connecting this node to another node above it (dubbed the parent node) is broken with some probability $q$, \textit{independent} of the number of edges incident to it\footnote{If there are no edges connected to this node from below, it is immaterial if there are subsequent edges going upward, since no particle can reach this node in the first place}. A quantity of particular interest is the probability that there exists an unbroken path from any of the leaves to the root. This can be solved straightforwardly owing to the recursive structure of the tree coupled with the lack of interactions between different edges incident on the node from below.

One might be tempted to cast the dynamics of entanglement on a tree in a similar light, with measurements and decoherence contributing to the severance of bonds with probability $p$ and $r$, respectively. However, such a simple picture fails to adequately capture the physics underlying the results we observe. The reason for this is twofold: firstly, each edge in bond DP exists in one of 2 states -- broken or unbroken. However, a quantum circuit can either transmit quantum information or only classical information, so these have to be represented by different bonds (such as an unbroken and ``partially" broken bond, respectively). Secondly, the probability of a bond emanating from a node to its parent \textit{depends on the state of the bonds entering it}, as suggested by the labeling of the transition matrix $W((a,b)\to c)$. Moreover, this probability depends on whether an input bond was broken by measurement or by decoherence (i.e.\ if the incident state was a product state or a maximally mixed state), even though a broken bond is incapable of carrying information (quantum or classical) all the same\footnote{For instance, when the edges incident on a root are in the c-states $s=2$ and $s=I$, the output states can be $s=2, s=1$ or $s=I$, while if the edges are $s=2$ and $s=\sigma$, the output states can only be $s=2$ (transmitting 1 qubit) and $s=\sigma$.}. Thus, each edge can exist in one of four inequivalent states, corresponding to the 4 c-states defined previously; $s=2$ corresponds to an unbroken bond capable of transmitting 1 qubit of information, $s=1$ to a ``partially broken" or imperfect bond that transmits only classical information, and $s=I, \sigma$ to completely broken bonds, albeit in different ways. Finally, such a mapping succeeds because no two bonds can interact to produce a bond that is ``less broken" than either of them. In order to produce an unbroken bond, at least one incident bond has to be in an $s=2$ state. Similarly, no number of broken bonds can conspire to create a (partially or completely) unbroken bond. This will be made explicit in subsequent sections once $W$ is calculated.

One can then obtain yet another physical picture for how information flows from the leaves to the root, especially by considering limiting cases. At each node of the tree, the input states $a$ and $b$ interact to produce an output state $c$ with probability $W((a,b)\to c)$. While ``transiting" between nodes, each edge can be subject to noise or measurement at rates $r$ and $p$ respectively, which causes this bond to ``break". When $p=r=0$, no bond is broken, so there exist multiple paths from the leaves to the tree. On the other hand, if $p=1$ or $r=1$, no path survives. Additionally, if $r\neq0$, paths from the root to a leaf can consist of both unbroken but imperfect bonds -- in this case, the path as a whole is imperfect in that it can only transmit classical information. The rates at which different types of bonds can emanate from a node (or the root itself) are encoded in the transition matrix $W$. The dynamics of information then reduce to the dynamics of $k^D$ particles, all starting at the leaves, as they attempt to traverse the tree. These particles can only move upward. Information is retained in the circuit only if at least one particle is able to reach the root. That information is classical or quantum, depending on the exact nature of the path available to these particles.  Two instances of such trees are presented in \cref{fig:GenDP}.

\section{``Standard" Measurement Induced Phase Transition}
\label{sec:mipt}

We first focus on the case where $r=0$. Our setup then emulates that of ``standard" measurement-induced phase transitions, where qubits are subject to unitary gates and projective measurements, but in the absence of decoherence or noise. Initially, each qubit is maximally entangled to a corresponding qubit in $L$, so the initial state is $P_0(2) = 1; P_0(s\neq2)=0$. In a tree circuit initially comprised only of maximally entangled $I=2$ states (for the leaves), projective measurements and unitary gates, the input to each node can only either be pure or maximally entangled, corresponding to the c-states $\sigma$ or $2$. The recursion relation \cref{eq:gen_rec}, previously defined on a 4-dimensional c-state space, now reduces to a recursion on the 2-dimensional space consisting of $\sigma$ and $2$ alone. By using the fact that $P_D(2) + P_D(\sigma)=1$, this reduces further to a single equation for $P_D(2)$, which is analyzed below.

As alluded to earlier, Clifford states only have discrete values of entropy. Therefore, the ensemble averaged mutual information is given by
\begin{equation}
    \overline{I}(R;L) = 2S(R) = 2 P_\infty(I=2).
\end{equation}

\begin{figure}
    \centering
    \includegraphics[width=0.45\textwidth]{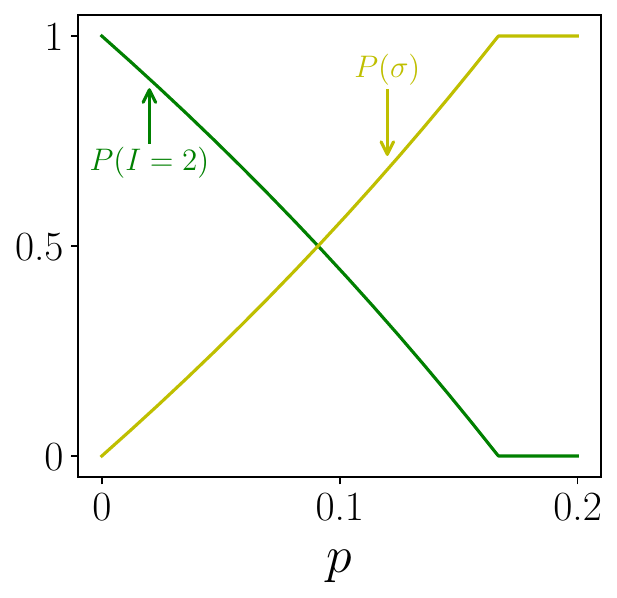}
    \caption{Probabilities of the mutual information $I(R;L)$ between the root and the leaves being 0 or 2, when the depth of the tree $D\to\infty$, as a function of the measurement rate $p$. Data are presented in the absence of measurements ($p=0$). The probability $P(\sigma)$ of the root having been disentangled from the leaves (``purified") steadily increases from $p=0$, before becoming 1 at $p_c=1/6$.}
    \label{fig:r0}
\end{figure}

To compute $P_D(2)$ with $D\to\infty$, we first determine the transition probability $W$.  When both inputs to a node are identical, the output is identical to the inputs as well, i.e. $W((a,a)\to a) = 1$, where $a \in \qty{2,\sigma}$. When the inputs are different, we define $\alpha$ via $W((\sigma,2)\to2) \equiv \alpha$ and $W((\sigma,2)\to \sigma) \equiv 1-\alpha$. Since we sample unitary gates uniformly from the 2-qubit Clifford group, the ensemble is invariant under single qubit rotations; the transition probabilities are independent of which specific product state the root is in, allowing us to label all product states by $\sigma$. These transition probabilities are schematically summarized in \cref{tab:rules_MIPT}.

\begin{figure}
    \centering
    \begin{tikzpicture}

	\draw (0,1) node[label={[label distance=0cm]above:$a$}]{} -- +(0,-0.5) arc[start angle = 360, end angle = 181,radius=0.5cm] -- +(0,0.5) node[label={[label distance=0cm]above:$L_a$}] {};
	
	\draw[dashed] (1,1) node[label={[label distance=0cm]above:$b$}]{} -- +(0,-1);

	\draw[->,line width=2] (1.5,0.75) -- node[pos=0.5] (O) {} +(2,0);
	
	\draw (5,1) -- +(0,-0.5) arc[start angle = 360, end angle = 181,radius=0.5cm] -- +(0,0.5 )node[label={[label distance=0cm]above:$L_a$}] {};
	
	\repCirc[0.5][($(O) + (-0.75,0.5)$)];
	
	\draw[->,line width=2] (0.5,-0.5) -- node[pos=0.5,label={[label distance=0cm]left:$\tr_{L_a}$}]{} +(0,-1.5);
	
	\begin{scope}[yshift=-4cm]
		\draw[red,dashed] (0,1) node[label={[label distance=0cm]above:$a$}]{} -- +(0,-1);
		
		\draw[dashed] (1,1) node[label={[label distance=0cm]above:$b$}]{} -- +(0,-1);

		\draw[->,line width=2] (1.5,0.75) -- node[pos=0.5] (O1) {} +(2,0);
		
		\draw[red,dashed] (4.5,1) -- +(0,-1);
		
		\repCirc[0.5][($(O1) + (-0.75,0.5)$)];
	\end{scope}
	
	\draw[->,line width=2,xshift=4cm] (0.5,-0.5) -- node[pos=0.5,label={[label distance=0cm]left:$\tr_{L_a}$}]{} +(0,-1.5);
	
\end{tikzpicture}
    \caption{A commutation diagram showing the equivalence between $W((M,\sigma)\to M)$ and $W((2,\sigma)\to 2)$. When qubit $a$ is in a mixed state, it is equivalent to a Bell pair between $a$ and another qubit $L_a$, represented by the vertical arrow on the left. $\mathbb{P}_aU$ is a unitary gate, followed by a projective measurement on $a$, that maps the pair of c-states $(2,\sigma)\to 2$. Starting from the top-left and proceeding clockwise, $\mathbb{P}_aU$ results in a Bell pair between $b$ and $L_a$. After tracing $L_a$ out, the maximally mixed state on $b$ remains. Crucially, neither $U$ nor $\mathbb{P}_a$ affect $L_a$, so $\comm{\tr_{L_a}}{\mathbb{P}_aU} = 0$, allowing us to traverse the diagram in another, equivalent way. Now moving \textit{counter-clockwise} from top-left, tracing $L_a$ out produces a mixed state $\mathbb{}$, and applying $\mathbb{P}_aU$ should nonetheless result in the same output $\mathbb{1}_b$. Therefore, each $\mathbb{P}_aU$ which has the effect $\mathbb{P}_aU:(2,\sigma)\to 2$ \textit{also} maps $(M,\sigma)\to\sigma$. Since $W$ is proportional to the number of such $U$, $W((M,\sigma)\to M) = W((2,\sigma)\to 2)$.}
    \label{fig:purif}
\end{figure}
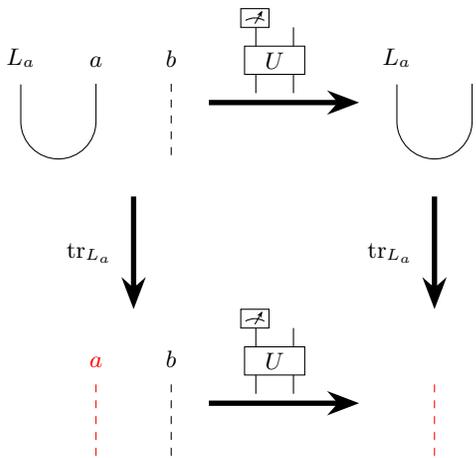

The calculation of $W((\sigma,2)\to2)$ can be simplified by utilizing the purification picture, shown in \cref{fig:purif}. A Bell pair between (the logical qubit of) the leaves and the root can be considered a purification of the \textit{maximally-mixed} state on the root, $\rho_R\propto\mathbb{1}$. The c-states $M$ and $I=2$ are then in a 1-to-1 correspondence, so that the transition probabilities $W((M,\sigma)\to M) \equiv W((2,\sigma)\to 2) = \alpha$, and $W((M,\sigma)\to \sigma) \equiv W((2,\sigma)\to \sigma) = 1-\alpha$.
Thus, we can focus on the calculation of $W((M,\sigma)\to M)$, which follows from simple combinatorial considerations. This equivalence is more explicitly demonstrated in \cref{app:M_2_equiv}.

     The aforementioned invariance under single-qubit rotations also allows us to assume that the measurement at the end of a unitary gate is done in the $Z$ basis alone, without losing generality. The stabilizer of the qubit at $a$ is taken to be $\sigma = X_1/Y_1/Z_1$ (corresponding to the c-state $\sigma$), while the qubit at $b$ is in the maximally mixed state $\rho_{b}\propto\mathbb{1}$ (corresponding to $M$). Under a Clifford 2-qubit gate, $\mathbb{1}\to\mathbb{1}$, while $\sigma$ is mapped to $\sigma'$, which can be one of 15 operators. When $\sigma'|_{1} = I$ or $Z$ -- but $\sigma'\neq Z_1I_2$ -- and qubit 1 is measured in the $Z$ basis, the resulting quantum state at the root is pure (the c-state is $\sigma$). This follows from the fact that $Z_1$ commutes with the stabilizers of the 2 qubit state, but is not in the stabilizer tableau. Measuring $Z_1$ then results in the tableau gaining a stabilizer. On average, this happens with a probability of 6/15 = 2/5. Otherwise, the c-state at the root is $M$. Thus, we have that $W((\sigma,M)\to \sigma) = W((M,\sigma)\to \sigma) = 2/5 \equiv 1-\alpha$ for random two-qubit Clifford gate with uniform distribution.

\begin{figure}    
    \begin{tikzpicture}
        \draw (0,0) -- (0,0.5);
        \meas[(-0.375,0.5)];
        \draw[dashed] (0,1) -- (0,1.5);
    \end{tikzpicture}\hspace{2em}
    \begin{tikzpicture}
        \draw[dashed] (0,0) -- (0,0.5);
        \meas[(-0.375,0.5)];
        \draw[dashed] (0,1) -- (0,1.5);
    \end{tikzpicture}\hspace{3em}
    \begin{tikzpicture}[scale=0.7]
        \draw (0,0) -- (0.75,0) node[right=0.2]{c-state $2$};
        \draw[yshift=-1cm,dashed] (0,0) -- (0.75,0) node[right=0.2]{c-state $\sigma$};
    \end{tikzpicture}
    \caption{Measurements on a maximally entangled ($I=2$) or a pure state ($\sigma$) both result in a pure state as the output. A measurement is independently made on each leg with probability $p$.}
    \label{fig:rules_meas}
\end{figure}
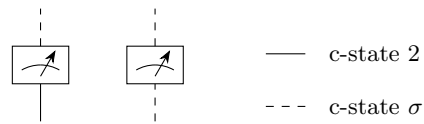

    Using the transition probability derived above, we now can formulate the recursion relation for $P_D(2)$. Following a measurement performed with probability $p$, the $P_D(2)\to\widetilde{P_D}(2) \equiv (1-p)P_D(2)$ (Cf. \cref{fig:rules_meas}). The Markov process on the c-state space can now be defined.

\begin{equation}
    \begin{aligned}
    P_{D+1}(2) &= \widetilde{P_D}(2)^2 + 2\alpha \widetilde{P_D}(2)(\widetilde{P_D}(\sigma))\\
    &= \widetilde{P_D}(2)^2 + 2\alpha \widetilde{P_D}(2)(1-\widetilde{P_D}(2)).
    \end{aligned}
    \label{eq:rec_mipt}
\end{equation}
The fixed points of this equation $P_\infty(s)$ are tabulated in \cref{tab:my_label}, while \cref{fig:r0} shows the values of $P(2)$ and the location of the entanglement phase transition, obtained upon solving \cref{eq:rec_mipt}. Above a critical measurement rate $p\geq p_c\equiv\frac{1}{6}$, the root is in a product state; its mutual information with leaves is 0 with probability 1. This transition is continuous, as expected.

\begin{table}[h]
    \begin{center}
        \begin{tabular}{|c| c|}
		\hline
		$W((a,b)\to c)$\newline&
		\adjustbox{margin=2mm}{
		\begin{tikzpicture}[baseline,scale=1]
			\pronglabel{(0,0)};
			\node[draw,circle,fill=white] at (0,0) {$U$};
		\end{tikzpicture}}\\
		\hline
		1&
		\adjustbox{margin=2mm}{\begin{tikzpicture}[baseline,scale=1]
			\prong{(0,0)}
			\node[draw,circle,fill=white] at (0,0) {$U$};
			
			\coordinate (O) at (3,0);
			\prong{(O)}[dashed]
			\node[draw,circle,fill=white] at (O) {$U$};
		\end{tikzpicture}}\\
		$\alpha$&
		\adjustbox{margin=2mm}{\begin{tikzpicture}[baseline,scale=1]
			\prong{(0,0)}[solid][solid][dashed]
			\node[draw,circle,fill=white] at (0,0) {$U$};
			
			\coordinate (O) at (3,0);
			\prong{(O)}[solid][dashed][solid]
			\node[draw,circle,fill=white] at (O) {$U$};
		\end{tikzpicture}}\\
		$1-\alpha$&
		\adjustbox{margin=2mm}{\begin{tikzpicture}[baseline,scale=1]
			\prong{(0,0)}[dashed][solid][dashed]
			\node[draw,circle,fill=white] at (0,0) {$U$};
			
			\coordinate (O) at (3,0);
			\prong{(O)}[dashed][dashed][solid]
			\node[draw,circle,fill=white] at (O) {$U$};
		\end{tikzpicture}}\\
		\hline
	\end{tabular}
    \end{center}
    \caption{The evolution rules for entanglement dynamics on the tree, obtained through the mapping to a classical Markov process. $U$ denotes a random 2-qubit Clifford unitary with one of the output qubits projectively measured.}
    \label{tab:rules_MIPT}
\end{table}

\begin{table}[h!]
    \centering
    \begin{tabular}{|c|c|c|}
    \hline
         $p<p_c$ & $P_\infty(M)>0$ & Stable  \\
         & $P_\infty(M)=0$ & Unstable\\
         \hline
         $p>p_c$ & $P_\infty(M)=0$ & Stable\\
         \hline
    \end{tabular}
    \caption{Fixed point solutions to \cref{eq:rec_mipt}, where $p_c$ is analytically determined to be 1/6.}
    \label{tab:my_label}
\end{table}

\begin{figure}
	\centering
	\includegraphics[width=0.45\textwidth]{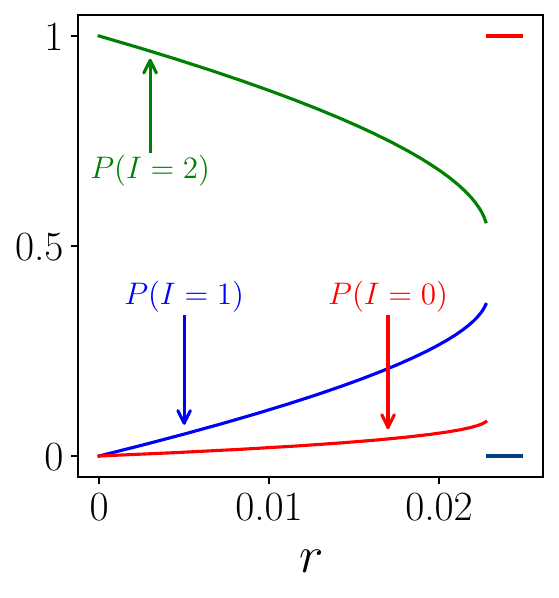}
	\caption{Probabilities of the mutual information $I(R;L)$ between the root and the leaves being 0, 1 or 2, when the depth of the tree $D\to\infty$, as a function of the decoherence rate $r$. Data are presented in the absence of measurements ($p=0$). The threshold values of $r$ below which the preservation of either classical or quantum information occurs are the same.}
	\label{fig:p0}
\end{figure}

\section{Noisy Entanglement Transitions}
\label{sec:noise}

We now turn our attention to cases where $r\neq0$. A 2-dimensional c-state space is no longer sufficient to describe these dynamics, since we need to distinguish between the qubits that are entangled to the leaves, and those that are entangled to the environment (i.e.\ maximally mixed; refer to \cref{sec:setup}). This precludes tracing the leaves out prior to the implementation of the circuit.

Pure states and mixed states ($\sigma$ and $M$ in the notation of \cref{sec:mipt}), while both being disentangled from the leaves, affect the dynamics in non-equivalent ways, and so have to be considered separately. Additionally, an interaction between $M$ and $I=2$ c-states can produce mixed states $\rho_{RL}$ with $I=1$. One such state could be $\rho_{RL} \propto \mathbb{1} + Z_R\prod\limits_{\qty{j}\subset L} Z_j$, where the root and the leaves are classically correlated.

\begin{figure}[!ht]
	\centering
	\includegraphics[width=0.45\textwidth]{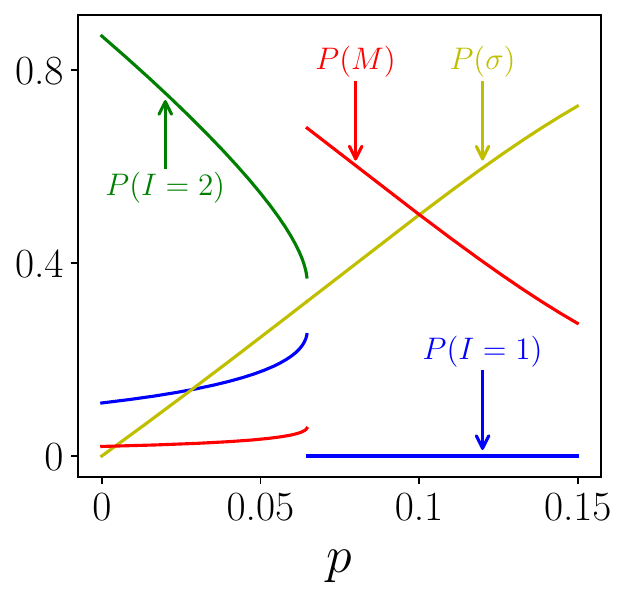}
	\caption{Probabilities of the mutual information $I(R;L)$ between the root and the leaves being 0, 1 or 2, when the depth of the tree $D\to\infty$, as a function of the measurement rate $p$, presented now in the presence of noise ($r=0.01$). The threshold values of $p$ below which the preservation of either classical or quantum information occurs are still identical. The transition has now changed from a second-order transition to a first-order one, where $P_\infty(I\neq0)$ abruptly jumps to 0.}
	\label{fig:r0.01}
\end{figure}

If the root is completely disentangled from the leaves, $\rho_{RL} = \rho_R \otimes \rho_L$, then $\rho_R \propto \mathbb{1}$ or $\rho_R \propto \mathbb{1} + \sigma$, for some Pauli operator $\sigma$. This corresponds to $P(I=0)\equiv P(M) + P(\sigma) = 1$. Consequently, we label the critical values of noise/measurement $r_c/p_c$ ``threshold" values, since above these critical rates (which depend on each other), no information is retained in the root. With the calculation of $W$ left to \cref{app:W_deriv}, the Markovian evolution on the full 4-dimensional c-space, consisting of the c-states $s\in\qty{2,1,\sigma,M}$, is

\begin{widetext}
\begin{equation}
    \begin{aligned}
P_{D+1}(2) &= \qty(\widetilde{P}_{D}(2))^2 + 2 \alpha \, \widetilde{P}_{D}(2) \widetilde{P}_{D}(1) + 2 \alpha \, \widetilde{P}_{D}(2) \widetilde{P}_{D}(\sigma) + 2\alpha\beta \, \widetilde{P}_{D}(2) \widetilde{P}_{D}(M) \\
P_{D+1}(1) &= \qty(\widetilde{P}_{D}(1))^2 + 2 \alpha \, \widetilde{P}_{D}(2) \widetilde{P}_{D}(1) + 2 \alpha \, \widetilde{P}_{D}(2) \widetilde{P}_{D}(M) + 2 \qty(\alpha\overline{\beta} + \overline{\alpha}\overline{\gamma}) \widetilde{P}_{D}(1) \widetilde{P}_{D}(M) \\
&\quad + 2 \alpha \, \widetilde{P}_{D}(1) \widetilde{P}_{D}(\sigma) \\
P_{D+1}(\sigma) &= \qty(\widetilde{P}_{D}(\sigma))^2 + 2 \alpha \, \widetilde{P}_{D}(\sigma) \widetilde{P}_{D}(2) + 2 \alpha \, \widetilde{P}_{D}(\sigma)\widetilde{P}_{D}(1) + 2 \alpha \, \widetilde{P}_{D}(\sigma) \widetilde{P}_{D}(M)\\
P_{D+1}(M) &= \qty(\widetilde{P}_{D}(M))^2 + 2 \alpha \, \widetilde{P}_{D}(\sigma) \widetilde{P}_{D}(M) + 2 \alpha \, \widetilde{P}_{D}(M) \widetilde{P}_{D}(1) + 2\overline{\alpha}\gamma \, \widetilde{P}_{D}(2) \widetilde{P}_{D}(M),
\end{aligned}
\label{eq:MEqn}
\end{equation}
\end{widetext}
where $\alpha,\beta,\gamma \in \qty[0,1]$ depend on the distribution from which the Clifford gates and the initial states are drawn, and $\overline{x}\equiv 1-x$ for $x=\alpha,\beta,\gamma$. When the gates are uniformly drawn from the Clifford distribution, their values are $\alpha_{\rm Clifford} = \frac{3}{5}, \beta_{\rm Clifford}=\frac{1}{3}$ and $\gamma_{\rm Clifford}=\frac{1}{2}$; this is the ensemble used in the remainder of this section. As before, $P_D(s)$ and $\widetilde{P}_D(s)$ respectively denote the probabilities of c-state $s$ before and after measurements and noise act on the root of a tree of depth $D$, that is, on the bottom and top ends of the noisy wires ${\red ||}$ in \cref{fig:schem1}.

\begin{equation}
    \begin{aligned}
    \widetilde{P}_D\qty(2) = (1-r)(1-p)P_D\qty(2)\\
    \widetilde{P}_D\qty(1) = (1-r)(1-p)P_D\qty(1)\\
    \widetilde{P}_D\qty(M) = r + (1-r)(1-p)P_D\qty(M)\\
    \widetilde{P}_D\qty(\sigma) = (1-r)\qty(p + (1-p)P_D\qty(\sigma))
    \end{aligned}
\end{equation}

A finite value of $P_\infty(2)$ indicates that, with finite probability, such circuits can be used to noise-resiliently encode one qubit of quantum information in the leaves. An important simplification arises when $p=0$. Since the initial state contains no product states, $\widetilde{P}_D\qty(\sigma)=0$ in the absence of measurements at all depths $D$. As $\sum\limits_s \widetilde{P}_D\qty(s) = 1$, it suffices to study an effectively two dimensional Markov equation for the c-states  involving only $P(2)$ and $P(M)$, since $P(1)$ can be replaced by $1-P(2)-P(M)$. The recursion relation can then be analytically solved. Below, we study the transition, now driven by noise, that is observed when the unitary gates are again drawn uniformly from the Clifford group. The asymptotic probabilities as functions of $r$ are shown in \cref{fig:p0}. For small values of $r$, $P(M) << 1$, indicating that quantum or classical information can be retrieved from the leaves with probabilities $P(I=2)$ and $P(I=1)$, respectively; both of these are nonzero. However, for $r>r_c\approx0.0225$, no information is retained, and the root is a maximally mixed state. Moreover, the noise threshold is identical for both classical and quantum information, meaning that the circuit fails to retain either classical or quantum correlations between $R$ and $L$ above the \textit{same} value of $r$.

Reintroducing measurements, the asymptotic values of $P(s)$ as functions of $p$, in the presence of noise $r=0.01$ are plotted in \cref{fig:r0.01}. This figure differs from \cref{fig:r0} in two ways. First, the critical rate $p_c$ has reduced with the introduction of noise. Second, the nature of the transition has itself changed, from a continuous to a discontinuous change in the values of $P(s)$ at $p_c$. This agrees with the predictions from the Ising model on the tree, with the transition remaining discontinuous for any value of $r>0$. 

\begin{figure}
    \centering
    \includegraphics[width=0.45\textwidth]{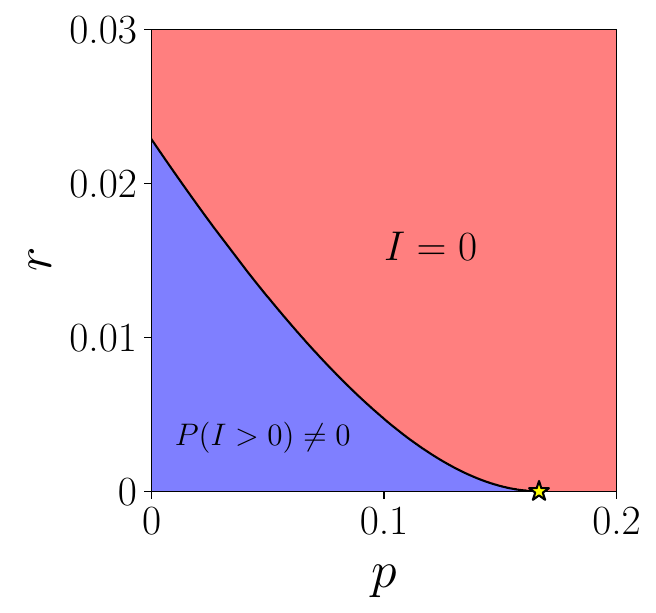}
    \caption{A phase diagram showing the noise and measurement thresholds for the successful preservation of quantum information between the root and the leaves. The black solid line denotes a phase boundary of first-order transitions (i.e.\ $P(I\neq0)$ drops abruptly to zero) for any $r>0$. In the absence of noise ($r=0$), the measurement-induced phase transition is a \textit{second} order transition. The critical measurement strength is analytically determined to be $p_c=\frac{1}{6}$, and is marked by \protect\tikz\protect\node[yshift=-50pt,star,star point ratio = 0.4,draw,fill=yellow!80!black,shape border rotate=180] at (1,1){}; on the $p$ axis. $P(I=2)$ and $P(I=1)$ vanish simultaneously across the phase boundary, indicating that the absence of a region where classical information can still be retained.}
	\label{fig:phasediag}
\end{figure}

The entire phase diagram, with $p$ and $r$ both varied, is shown in \cref{fig:phasediag}. The blue region denotes the values of $r$ and $p$ where information stored in the leaves \textit{can} be retrieved from the root, while the red region denotes the measurement/noise rates for which \textit{no} information persists. The phase boundary is a curve across which  $P(I=0)\equiv P(M)+P(\sigma)$ jumps discontinuously, except at the noiseless point $r=0, p=1/6$. Note that in the red region, the root can either be a mixed or a pure state, but it is nonetheless disentangled from the leaves. This family of circuits has \textit{identical} thresholds for both classical and quantum information, i.e. there is no region in the phase diagram where classical information is preserved, but quantum information is not.

\section{Extending the Phase Diagram}
\label{sec:ext}

\subsection{Non-uniform Gate Ensembles}
\label{ssec:nonunif}
A natural question arising from the derivation in (appendix) concerns the role of the parameters $\alpha, \beta, \gamma$ in \cref{eq:MEqn}. In results presented thus far, $P(2/1)$ seem to vanish simultaneously, suggesting that the noise thresholds for quantum and classical information are the same. However, there is no fundamental obstruction to there being different noise thresholds; in fact, in practical settings, classical information is often easier to store in a noise-resilient fashion. In pursuit of a regime where $P_\infty(2)=0$ but $P_\infty(1)\neq0$, we study how the phase diagram evolves as $\alpha, \beta, \gamma$ are varied. For concreteness, we focus on the case where $p=0$, since measurements are not expected to provide qualitatively new phenomena. These parameters can be varied by changing the distribution from which the unitary gates are drawn. We present one such method for tuning $\alpha$, while emphasizing that such a method is one of many.

\begin{figure}[!h]
    \raggedright
    \vspace{2em}
    \begin{tikzpicture}[scale=1]
		
		\foreach \x in {0,1cm}
		{
			\draw[xshift=\x] (0,-0.5) -- (0,2);
		}
		
		\draw[fill=white] (-0.3,0.5) rectangle node {$CZ$} (1.3,1);
		\node[rectangle,draw,fill=white] at (0,1.5) {$H$};
            \node[draw,fill=white] at (0,0) {$H$};
		\meas[(-0.4,2)]
		
		\begin{scope}[xshift=3cm]
			\foreach \x in {0,1cm}
			{
				\draw[xshift=\x] (0,-0.5) -- (0,2);
			}
			
			\draw[fill=white] (-0.3,0.5) rectangle node {$CZ$} (1.3,1);
			\node[draw,fill=white] at (0,1.5) {$H$};
			\node[draw,fill=white] at (1,0) {$H$};
			\meas[(-0.4,2)]
		\end{scope}

        \begin{scope}[xshift=5cm]
	   \meas[(0,1)]\node at (2.1,1.25) {$Z$ Measurement};
        \end{scope}
	\end{tikzpicture}
    \caption{Two unitary gates with distinct values of $\alpha$. The gate on the left (right) has $\alpha = 0 (1)$. By sampling from this two-member set with different biases, the desired value of $\alpha$ can be obtained.}
    \label{fig:tuningA}
\end{figure}
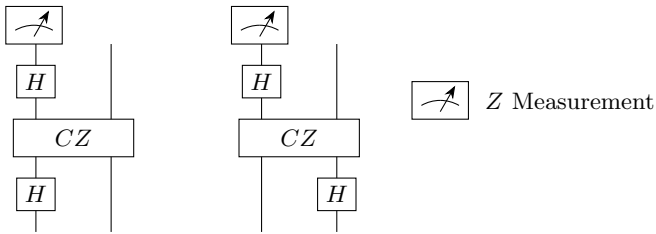

Consider the two gates shown in \cref{fig:tuningA}. Gate 1 has $\alpha=0$, while gate 2 has $\alpha=1$. These calculations are shown in \cref{app:a0a1gates}. If the Clifford gates are chosen to be Gate 1 with an arbitrary probability $\alpha'$, and Gate 2 with probability $1-\alpha'$, this ensemble will result in an $\alpha=\alpha'$. Both gates have $\beta=\gamma=0$, so the resulting $\beta$ and $\gamma$ of the ensemble will be 0, too. It should be pointed out that this ensemble is no longer invariant under single qubit Clifford rotations, so in addition to specifying the gates, one should also specify the initial conditions. We elaborate in \cref{app:a0a1gates} that our method is nonetheless applicable, allowing us to determine noise thresholds as before.

\begin{figure}
	\raggedright
	\hspace{-2em}
	\includegraphics[width=0.475\textwidth]{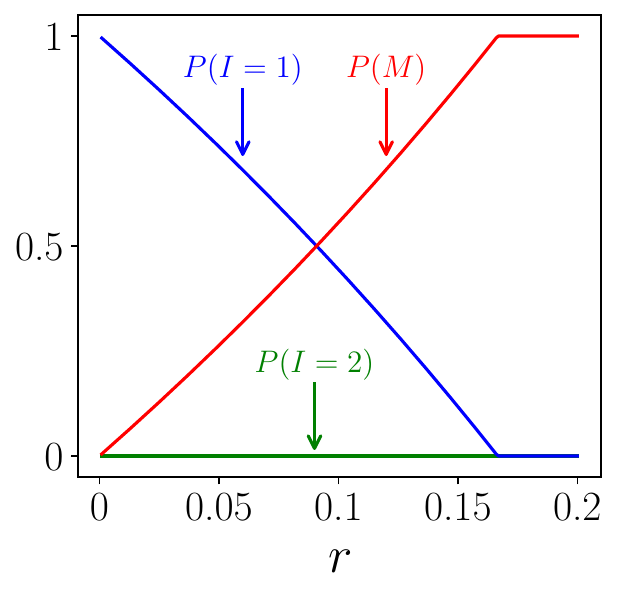}
	\caption{The noise-induced phase transition in the probability $P(I=1)$ that the tree quantum circuit can transmit classical information, in the absence of measurements ($p=0$) and with  $\alpha=\frac{2}{5}=1-\alpha_{\rm Clifford}, \beta = \frac{1}{3}$ and $\gamma=\frac{1}{2}$. Under a symmetry transformation (see text below), this problem can be mapped to the study of the MIPT of \cref{sec:mipt} with $\alpha\xleftrightarrow{}1-\alpha$. The noise thresholds are $r_c(2)=0$ and $r_c(1)=1/6$.}
    \label{fig:a0.4}
\end{figure}

To demonstrate one effect of tuning the gate ensembles, we first fix $\beta=\beta_{\rm Clifford}$ and $\gamma=\gamma_{\rm Clifford}$. We find that there are 3 distinct regimes, determined by whether $\alpha <,= $ or $>0.5$. Let $r_c(s)$ denote the noise thresholds for quantum ($s=2$) and classical ($s=1$) information. $r_c(s)$ is defined as the smallest value of $r$ for which $P_\infty(s)=0$, beginning from an initial state where the qubits form Bell pairs with the leaves. The case of $\alpha=0.6>0.5$ has already been studied in the previous section, so we now turn to $\alpha=0.4<0.5$. Here, $P(I=2)$ goes to zero for \textit{any} finite $r>0$, showing that this circuit cannot carry quantum information; its noise threshold $r_c(2)=0$. However, $r_c(1)>0$, indicating that it can still transmit classical information, robustly under noise. This transition is shown in \cref{fig:a0.4}. One might notice that \cref{fig:a0.4} looks remarkably similar to \cref{fig:r0}. This stems from the fact that \cref{eq:gen_rec} is symmetric under the transformations $P(I=2)\xleftrightarrow{} P(I=1)$, $P(\sigma)\xleftrightarrow{} P(r)$, $\alpha\xleftrightarrow{}1-\alpha$ and $p\xleftrightarrow{} r$. Therefore, the noisy transition is a continuous one, akin to the MIPT observed in \cref{fig:p0}. Moreover, since $\alpha=0.4 = 1-\alpha_{\rm Clifford}$, the critical noise rate $r_c(1) = 1/6$, identical to that of the $p_c$ for the MIPT. We have also confirmed that this phenomenon is robust across a wide range of $\beta {\rm\ and\ } \gamma$.

\begin{table}[h]
    \centering
    \begin{tabular}{|c|c|c|}
    \hline
         $\alpha<0.5$ &$r_c(2)=0$ & $r_c(1)\neq0$\\
         \hline
         $\alpha>0.5$ & $r_c(2)\neq0$ & $r_c(1) = r_c(2)$\\
         \hline
    \end{tabular}
    \caption{Two of three qualitatively different behaviors of the noise thresholds, obtained in regimes accessed by tuning $\alpha$.}
    \label{tab:alpha}
\end{table}

\subsection{Boundary Fields}
\label{ssec:b_pert}

\begin{figure}
	\raggedright
	\hspace{-2em}
	\includegraphics[width=0.475\textwidth]{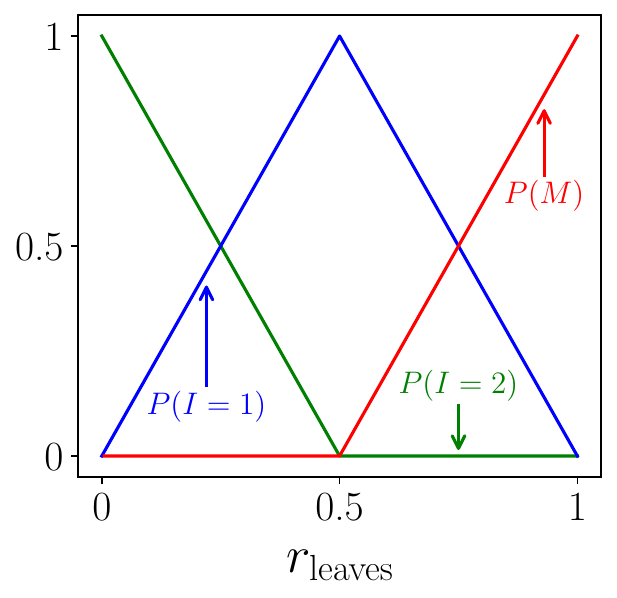}
	
	\caption{The continuous transitions in the probabilities that the circuit can transmit information, driven now by decoherence acting \textit{solely} on the leaves at a rate $r_{\rm leaves}$, again in the absence of measurements ($p=0$). Quantum information cannot be transmitted for $r_{\rm leaves}>0.5$, while classical information can be transmitted for some finite probability that $r<1$. The thresholds are $r^c_{\rm leaves}(2)=0.5$ and $r^c_{\rm leaves}(1)=1$}.
    \label{fig:a0.5_r0}
\end{figure}

The self-dual point $\alpha=0.5$ deserves special attention. Here, we find that \textit{any} finite rate of noise corrupts the information in the circuit, so $r_c(1,2) = 0$. However, one still observes transitions when the \textit{leaves alone are subject to decoherence}. The restriction our attention to boundary noise is motivated by the results on the Ising tree, which were shown in \cref{fig:IsingMI} (right). Moreover, the noise thresholds are now \textit{different} for the transmission of classical or quantum information. To distinguish between the two rates of noise, we term the boundary noise rate $r_{\rm leaves}$ and label its corresponding thresholds as $r^c_{\rm leaves}(s)$. $P(I=2)$ undergoes a continuous phase transition at $r^c_{\rm leaves}(2)=0.5$, above which $P(I=2)=0$. There remains a finite probability that classical information is transmitted for $r<1$, so $r^c_{\rm leaves}(1) = 1$, although $P(I=1)$ encounters a nonanalyticity at $r = r^c_{\rm leaves}(2)$. These results, with $\beta=\beta_{\rm Clifford}$ and $\gamma=\gamma_{\rm Clifford}$ as in the previous subsection, are presented in \cref{fig:a0.5_r0}.

The introduction of boundary noise corresponds to choosing the initial conditions to \cref{eq:MEqn}. If the boundary noise rate is $r_{\rm leaves}$, the persistence of information to the root is determined by the solutions to \cref{eq:MEqn} with $r=0$, with the initial conditions $P_0(2)=1-r_{\rm leaves}, P_0(M) = r_{\rm leaves}$ and $P_0(1)=P_0(\sigma)=0$.

\subsection{Multi-step Protocols}
\label{ssec:mulstep}
\begin{figure*}
    \centering
    \includegraphics[width=\textwidth]{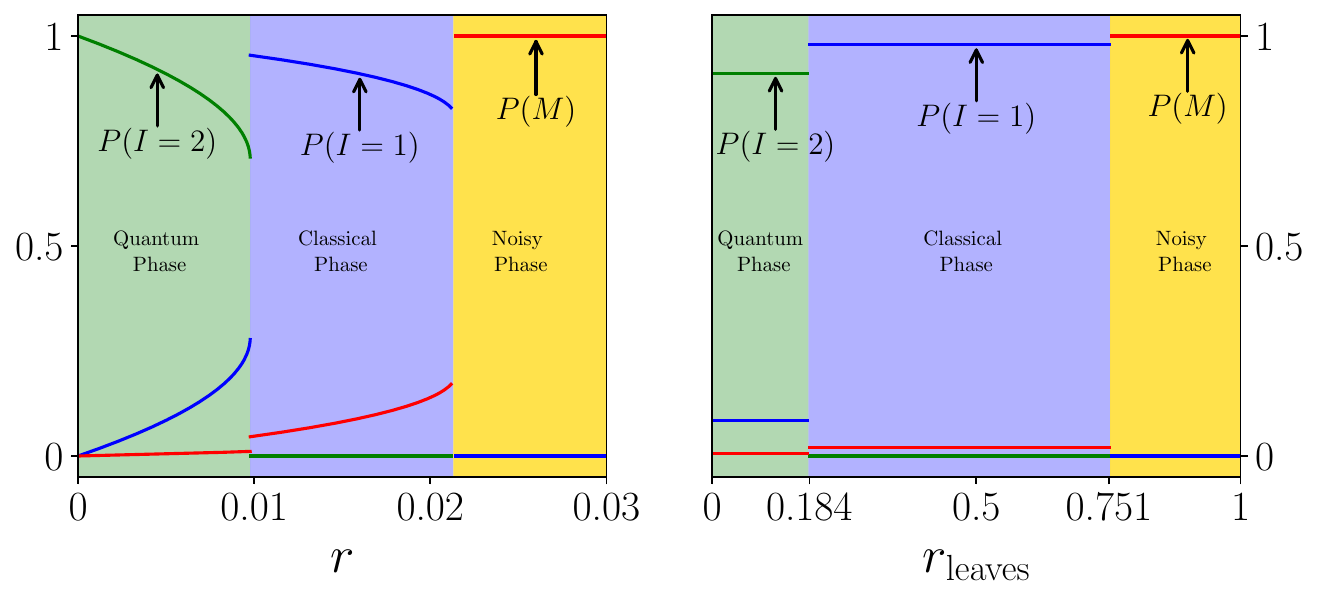}
	\caption{Information transmission capacity transitions driven by decoherence acting on the (left) entire circuit and (right) leaves, in two-step tree quantum circuits. In both cases, the transitions are discontinuous/first-order. The circuits exhibit three distinct phases -- the quantum, classical and noisy phases, where quantum, classical and no information are transmitted, respectively. The parameters for the unitary gates acting in the even/odd layers are $\alpha_{e/o}=0.8/0.2, \beta_{e/o}=\gamma_{e/o}=0$. (Right) The transition driven by the boundary decoherence rate $r_{\rm leaves}$ in the presence of global noise $r=0.005$. Note that the transmission probabilities $P(s)$ for $s=2,1,M$ are piecewise constant in their respective phases, unlike the results shown in the left panel. The noise thresholds $r^c_{\rm leaves}(2)\approx0.184$ and $r^c_{\rm leaves}(1)\approx0.751$ are marked on the x-axis.}
    \label{fig:mulstep}
\end{figure*}

Yet more exotic phenomena are observed in set ups where the gate ensembles vary with depth $D$, which has hitherto, not been the case. We choose the simplest deviation -- one ensemble for $D$ even, identified by $\alpha_e$, and one for $D$ odd, identified by $\alpha_o, $. While we set $\beta=\gamma=0$ across the all layers (so that this $D$-dependent ensemble can be perfectly emulated using the gates from \cref{fig:tuningA}), identical results are observed for $\beta,\gamma \neq 0$. For definitiveness, we focus on the case without measurements, $p=0$, although the extension is straightforward.

When $\overline{\alpha} = (\alpha_e + \alpha_o)/2 = 0.5$, we discover a rich phase diagram, which can be traversed by \textit{independently} tuning (a) the global noise rate $r$ throughout the entire quantum circuit , and (b) the leaves' noise rates ($r_{\rm leaves}$). The phase diagram contains 3 \textit{distinct} regions -- one where quantum information can be transmitted, one where only classical information can be transmitted, and one where no information is transmitted, from the leaves to the roots. The noise thresholds are finite \textit{and} different, i.e.\ $r^c(2)\neq r^c(1)$, and $r^c_{\rm leaves}(2;r)\neq r^c_{\rm leaves}(1;r)$\footnote{The definitions of the leaf noise thresholds have been extended to incorporate the effect of the global noise $r$. $r^c_{\rm leaves}(2/1;r)$ is the largest rate of boundary decoherence $r_{\rm leaves}$ at which quantum/classical information can be transmitted from the leaves to the root with nonzero probability, for a fixed rate $r<r^c(2)$ of noise on the entire circuit.}.

\subsubsection{Tuning Global Noise $r$}

As $r$ is tuned from zero, there is a phase where quantum information survives the circuit with a nonzero probability $P(2)$, albeit $P(2)$ decreases continuously as $r\to r^-_c(2)$. For $r<r_c(2)$, the failure rate\footnote{Recall that the failure rate is $P(I=0)\equiv P(M) + P(\sigma)$, but $P(\sigma)=0$ when $p=0$.} is small, thus \textit{some information}, either classical or quantum, almost always reaches the root.

At $r= r_c(2)\sim 0.01, P(2)=0$ drops to zero discontinuously, indicating that only classical information can persist. Even here, $P(1)$ decreases continuously $r\to r^-_c(1)$. On the other hand, $P(0)$ increases abruptly at $r=r_c(2)$, thus in fraction of realizations equal to $P(0)$, no information persists. Above $r_c(1)$, the failure rate is 1, and the circuit is unable to protect information from noise.

\subsubsection{Tuning Leaf Noise $r_{\rm leaves}$}

Three similar phases are observed if $r=0$ is fixed and $r_{\rm leaves}$ is tuned instead. However, the asymptotic values of $P(I)$ are now piecewise continuous across the phases, instead of varying continuously with $r_{\rm leaves}$. The results from the mean field calculation (\cref{fig:IsingMI}, Right) capture some aspects of this transition, while the true quantum model alone exhibits three distinct phases.

Perhaps most strikingly, the transition driven by boundary noise persists even in the presence of global noise $r > 0$ (unlike in the single-step setups of \cref{ssec:b_pert}). In the two non-noisy phases obtained by tuning $r_{\rm leaves}$, quantum (classical) information is still recovered with the same probability throughout the quantum (classical) phase. These results are encapsulated in \cref{fig:mulstep}.

It therefore appears that for the ensembles of tree circuits considered here, one requires at least a two-step protocol in order to observe distinct thresholds for quantum and classical information. Moreover, noise thresholds are lower for a two-step protocol than for a uniform protocol.

\section{Conclusion}
\label{sec:conc}

In this work, we studied the dynamics of quantum information transmission in families of noisy quantum circuits defined on trees, as quantified by the mutual information between the root and the leaves of the tree. Moreover, the model we considered is analytically tractable, allowing us to obtain exact results for noise and measurement thresholds. In the absence of noise, we were able to obtain an exactly solvable model of measurement induced phase transitions in tree circuits. While not the first such model, the benefit of our model over the one studied in \cite{feng2023tree} is that the calculation is far simpler, can be done without approximations, and lends itself to a physical interpretation in terms of a directed percolation process.

Upon the introduction of noise, we found that our model could still be exactly solved -- analytically or numerically -- allowing us to uncover a model of the MIPT which is robust to noise. This is an important step in the experimental realization of the MIPT, which is erased by noise in conventional 1+1D brickwork quantum circuits. In its own right, the system exhibits a panoply of phase transitions in its ability to resiliently transmit information from roots to the trees. Moreover, as is expected from statistical mechanics models on trees, our analyses revealed qualitatively different phase transitions driven independently by global and boundary decoherence. We showed that the central quantity of interest is the transition matrix $W$ and its associated recursion relation, and explored a portion of its rich landscape.

Lastly, we comment on the method itself developed in this work to tackle this problem. Since Clifford circuits are easy to simulate, one might naturally wonder at the absence of direct Clifford simulations of the tree circuit. However, we found that a recursive implementation of the circuits could only approach depths of $D\sim30$, which are \textit{far} too small to provide concrete evidence for the phase transitions studied in this paper. Even though each node could be simulated in $t = O(1)$ time (with a small prefactor), an algorithm to study a tree of depth $D$ with degree $k+1$ would take $O(tk^D)$ steps; the exponential growth is unavoidable here. While one could consider bootstrap methods as in \cite{feng2023tree}, we find that our method obviates the need for any direct simulation of the tree. It is our hope that the model can be analyzed in greater detail by utilizing the machinery of these recursion relations, and expect that it could find utility in a study of related problems on trees or tree-like circuits.

\begin{acknowledgments}
	We gratefully acknowledge computing resources from Research Services at Boston College and the assistance provided by Eliot Heinrich in optimizing large-scale direct Clifford simulations of tree circuits. We are especially thankful to Yaodong Li for several insightful discussions, and for directing us to previous studies on trees relevant to our work. This work also benefited from discussions with Ethan Lake at its incipience. This research is supported in part by the National Science Foundation under Grant No. DMR-2219735 (V. R. and X. C.). This research was also supported in part by the Department of Energy under Grant No. DE-SC0024324 (Y. H.), and the Air Force Office of Scientific Research under Grant FA9550-24-1-0120 (Y. H.).
\end{acknowledgments}

\appendix

\section{Determining Effective Magnetic Fields in Tree Ising Models}
\label{app:treeising}

For completeness, we briefly present the process by which the magnetic field at the root can be obtained \cite{lucassm7}. Assuming uniform Ising couplings between any two spins $i$ and $j$ joined by an edge (the pair is denoted $\left\langle i,j\right\rangle$), the Hamiltonian can be written as

\begin{equation*}
        H = -\sum\limits_{\left\langle i,j\right\rangle} s_is_j - \sum\limits_i h_{i} s_i,
\end{equation*}
where $\qty{s}$ are Ising variables which can take values $\pm1$. The recursive structure of the tree allows us to rewrite the Hamiltonian in a more illuminating form

\begin{equation}
     H = - \sum\limits_{D} \sum\limits_{i\in \qty{s_D}} \sum\limits_{j\to i} s_j(s_i + h_{D}),
\end{equation}
where $\qty{s_D}$ refers to the spins at depth $D$ (as measured from the base of the tree), and $j\to i$ is short-hand for ``$j$ is a child\footnote{``Child" has the following definition -- for a spin $i$ at depth $D$, its children are the $N_{\rm br}$ spins at depth $D+1$ that are connected to spin $i$. $N_{\rm br}$ is the number of branches at each node, which is constant in the regular trees considered in this work} of $i$". Moreover, we have focused now on a magnetic field $h_0$ that can vary with depth, but remains constant at a given depth.

Since the children of a given node only interact through their parent node, the Hamiltonian can be decomposed into separate Hamiltonians $H_D = \sum\limits_{i\in \qty{s_D}} \sum\limits_{j\to i} s_j(s_i + h_{D})$ at each depth $D$. $H_D$ only consists of couplings between a spin at depth $D$ and another at depth $D-1$; there are no couplings between spins at the same depth. Therefore, beginning from the putative ``bottom" of the tree, one can integrate these spins out to obtain an effective magnetic field $h_R$ on their parent spins (See \cref{fig:tree_solve_schem}). Since any regular tree of depth $D$ can be constructed from $N_{\rm br}$ trees of depth $D-1$, this procedure proceeds iteratively, starting from a tree of depth $D=2$. The calculation of the partition function $Z = \sum\limits_{s} e^{-\beta H}$ inherits this recursive simplification, as we now show.

Consider a tree of depth 2. The parent (i.e.\ root) spin is represented by $S_R$, and its children by $\qty{s_j}_{j=1}^{N_{\rm br}}$. The Hamiltonian
\[H = -\sum\limits_{j=1}^{N_{\rm br}} s_j(S_R + h_{0}) - h_{1}S_R\]
visibly factorizes, such that $Z$ can be written as

\begin{equation}
\begin{aligned}
    Z &= \sum\limits_{S_R} e^{\beta h_{1} S_R}\prod\limits_j \qty(\sum\limits_{s_j} e^{\beta s_j (S_R+h_{0})})\\
    &= \sum\limits_{S_R}e^{\beta h_{1} S_R}(2\cosh(\beta\qty(S_R+h_{0})))^{N_{\rm br}}\\
    &\equiv A \sum\limits_{S_R} e^{\beta (h_R)S_R},
\end{aligned}
\label{eq:Z_2spins}
\end{equation}
where $A$ and $h_R$ are obtained by equating the expressions in the second and third lines of \cref{eq:Z_2spins}, using the identity provided in \cite{lucassm7}
\[\frac{\cosh(\beta (h_0 + 1))}{\cosh(\beta (h_0 - 1))} = \exp\qty(2 \tanh^{-1}\qty(\tanh(\beta h_0)\tanh(\beta))).\]

As noted in \cref{eq:def_DH}, we are interested in the magnetic field alone, so we focus on $h_R$, excluding $A$. $h_R$ -- relabelled as $h_{R,1}$ to make the depth dependence explicit -- is given by

\[\tanh\qty(\beta\frac{h_{R,1}-h_{1}}{N_{\rm br}}) = \tanh\qty(\beta h_{R,0})\tanh\qty(\beta).\]

$h_{R,0}$, the magnetic field on the leaves of the trees, is equal to $h_{0}$, since there are no subsequent spins to create an effective field. However, the effect of the children spins can be represented wholly as a magnetic field on the root alone. We have shown how this is done for one level of spins; this process can now be extended to general $D$ by replacing $h_0$ by $h_{R,D}$, $h_{R,1}$ by $h_{R,D+1}$ and $h_1$ by $h_{D+1}$. Gathering $N_{\rm br}$ such spins leads to the general recursion relation

\begin{equation}
    \tanh\qty(\beta\frac{h_{R,D+1}-h_{0,D+1}}{N_{\rm br}}) = \tanh\qty(\beta h_{R,D})\tanh\qty(\beta),
    \label{eq:rec_ising}
\end{equation}

where $h_{R,D}$ denotes the magnetic field at the root of a tree with depth $D$\footnote{By convention, the leaves are chosen to be at depth 0.}, and $N_{\rm br}$ is the number of branches connected to this root. The boundary conditions on the leaves set the initial conditions $h_{R,0}\equiv h_{0}$; specifically, $\sigma_L = \uparrow/\downarrow \iff h_{R,D=0} =\pm\infty $. This is a  generic feature of circuits defined on a tree -- bulk properties are reflected in the parameters of the recursion relation, while boundary conditions are encoded in the initial conditions.

In the absence of a bulk magnetic field, $h_{D\neq0} = 0$, boundary spins of opposite orientations lead to exactly opposite magnetic fields at the root, straightforwardly recovering the mean field calculation for the MIPT.

\section{Equivalence between c-states $M$ and $2$ in the Absence of Noise}
	\label{app:M_2_equiv}

	We showed in \cref{sec:mipt} that when the inputs to a node are different, the explicit calculation of $W$ can be simplified since $W((\sigma,2)\to2/\sigma) = W((\sigma,M)\to M/\sigma)$. If a Bell pair between leg $a$ and $L$, and a pure state $\sigma$ on leg $b$ meet at a node, they will produce a Bell pair between the new root $c$ and $L$ with a probability $W((2,\sigma)\to2)$. In this case, tracing $L$ out turns the output state into a maximally mixed state. The output state $c$ can also be a pure state with probability $W((2,\sigma)\to\sigma)=1-\alpha$, in which case it remains a pure, product state supported on $R$, even if $L$ is traced out. This can also be shown directly from the evolution rule \cref{eq:dyn_rule} in the main text.
    
    Let $\rho_i = \rho_{a}(\sigma)\otimes\rho_{bL}(2)$ be the joint density matrix of the two input nodes, where $\rho_{bL}(2)$ corresponds to a state on qubit $b$ that is maximally entangled with the leaves, and $\rho_{a}(\sigma)$ is a pure state on qubit $a$. Evolving $\rho_i$ under \cref{eq:dyn_rule} only results in a state $\rho'_{b,L}(s)$ that is either pure or maximally entangled to $L$, again (i.e.\ $s=2$ or $s=\sigma$). Moreover, the output state $s$ depends on the specific unitary gate $U$ alone. Since the gates are drawn from a distribution, $\alpha$ denotes the probability that $s=2$ (that is, $\alpha \equiv W((2,\sigma)\to2)$).
	
	We begin by considering a specific gate $U$ such that the output is $\rho'(2)$, supported on $b$ and $L$. Upon tracing $L$ out, we obtain a maximally mixed state
	\begin{equation}
		\Tr_L\qty[\rho'_{bL}(2)] \propto \mathbb{1}_b,
	\end{equation}
	as can be seen from explicitly tracing the logical qubit $L'$ out in \cref{eq:eg_I2}. However, 
	
	\begin{equation}
		\begin{aligned}
			\Tr_L(\rho'_{bL}(2)) &\propto \Tr_L\qty[\Tr_a\qty[\mathbb{P}_a U_{ab} \rho_i U^\dagger_{ab} \mathbb{P}_a]]\\
			&=\Tr_a\qty[\Tr_L\qty[\mathbb{P}_a U_{ab} \rho_i U^\dagger_{ab} \mathbb{P}_a]]\\
			&=\Tr_a\qty[\mathbb{P}_a U_{ab} \qty(\rho_a(\sigma)\otimes\Tr_L\qty[\rho_{bL}(2)]) U^\dagger_{ab} \mathbb{P}_a]\\
			&=\Tr_a\qty[\mathbb{P}_a U_{ab} \qty(\rho_a(\sigma)\otimes\mathbb{1}) U^\dagger_{ab} \mathbb{P}_a].
		\end{aligned}
	\end{equation}
	
	Thus, a unitary gate $U$ that evolves $\rho(2)\otimes\rho(\sigma) \to \rho(2)$ \textit{also} evolves $\mathbb{1}\otimes\rho(\sigma) \to \mathbb{1}$ (and vice versa). Since $\alpha$ is proportional to the number of such gates, we can determine $\alpha$ by counting the number of gates that send $\mathbb(1)\otimes\rho(\sigma) \to \mathbb(1)$ under the evolution given by \cref{eq:dyn_rule}. Treating the ancillas as a purification of a mixed state will be a crucial ingredient in calculating $W((a,b)\to c)$ in more general cases, and their utility is demonstrated in \cref{app:W_deriv}.

\section{Derivation of the transition Matrix $W$}
\label{app:W_deriv}

In this appendix, we discuss the explicit calculation of $W$ in the 4-dimensional c-state space, ultimately resulting in the master equation \cref{eq:MEqn}. Specifically, we show how the transition probabilities can all be related to the transition rates for $W((2,1)\to c)$ by using purifications and partial traces, in the same vein as the derivation of $W$ for the standard MIPT (\cref{sec:mipt}).

	We begin by considering a node\footnote{Recall that each node in a tree consists of a 2-qubit unitary gate and a projective measurement of one of the qubits, taken here to be $a$. The ensembles of unitary gates that we consider are all invariant under the exchange of $a\leftrightarrow b$, so this choice is ultimately one of convenience.} of the tree whose inputs are $\rho_{a,L_a}(2)$ and $\rho_{b,L_b}(1)$. These states can be interpreted as the roots of two independent trees, where the state $\rho_{a,L_a}(2)$ is a Bell pair between the root $a$ and its logical leaf qubit $L_a$, while $\rho_{b,L_b}(1)$ is a mixed-state with a mutual information of 1 (i.e. only classical correlations) between $b$ and the corresponding logical leaf qubit $L_b$. As mentioned in the text, a representative example is $\rho_{b,L_b}(1)\propto \mathbb{1} + Z_b Z_{L_b}$. Although the leaves of the tree following the action of this node will nominally be $L_a\cup L_b$, one can again determine a logical qubit in this new set of leaves $L$ which will capture all the correlations between the output qubit $b$ and the leaves.
	
	In general, $W((a,a)\to a)=1$,  $\forall a \in \qty{2,1,\sigma,M}$. We therefore turn to the case where the input c-states are different, specifically, 2 and 1. When $\rho(2)$ and $\rho(1)$ interact at a node, the output is only either $\rho_{b,L}(2)$ (with probability $\alpha$, say) or $\rho_{b,L}(1)$ (with probability $1-\alpha$). However, not all $\rho_{b,L}(2)$ are equivalent -- with probability $\beta$, $\rho_{b,L}(2)$ is supported on $b$ and $L_a$ alone, while it includes $L_b$ with a probability $1-\beta$; a stabilizer state with the stabilizers $X_b X_{L_a}$ and $Z_b Z_{L_a} X_{L_b}$ is one such state\footnote{However, this can be transformed into a state with support on $L_a$ and $b$ alone by unitary gates that act exclusively on the leaves, so both states have $I=2$.}. Likewise, with probability $\gamma$ or $(1-\gamma)$, $\rho_{b,L}(1)$ is either supported on $b\cup L_b$ alone, or on $b\cup L_b\cup L_a$. These possibilities, along with the stabilizers of examples states, are summarized in a Markov tree, where the labels above the edges denote the probabilities for those transitions, and $\bar{x}\equiv 1-x$ for $x=\alpha,\beta,\gamma$.\\
	
    \begin{figure}
        \begin{tikzpicture}[
		level 1/.style = {level distance=2cm, sibling distance=4cm},
		level 2/.style = {sibling distance=2.5cm, level distance=2.5cm}]
		\node (R) {$\rho_{a,L_a}(2)\otimes\rho_{b,L_b}(1)$}
		child { node (I2) {$I=2$} 
				child { node[align=center] (I2a) {$\rho_{b,L_a}(2)$\\[1em]
				$X_bX_{L_a}$\\
				$Z_bZ_{L_a}$}
			edge from parent [->] node [pos=0.7,label=above:$\beta$] {}}
				child { node[align=center] (I2b) {$\rho_{b,L}(2)$\\[1em]
						$X_bX_{L_a}$\\
						$Z_bZ_{L_a}Z_{L_b}$}
					edge from parent [->] node [pos=0.7,label=above:$\overline{\beta}$] {}}
				edge from parent [->] node [pos=0.7,label=above:$\alpha$] {}
			}	
		child {node (I1) {$I=1$}
			child { node[align=center] (I1a) {$\rho_{b,L_b}(1)$\\[1em]
					$X_bX_{L_b}$\\
					}
				edge from parent [->] node [pos=0.7,label=above:$\gamma$] {}}
			child { node[align=center] (I1b) {$\rho_{b,L}(1)$\\[1em]
					$X_bX_{L_b}$\\
					$X_bZ_{L_a}$}
				edge from parent [->] node [pos=0.7,label=above:$\overline{\gamma}$] {}}
				edge from parent [->] node [pos=0.7,label=above:$\overline{\alpha}$] {}
		};		
	\end{tikzpicture}
    \caption{A Markov tree representing the four different families of states that can result from the c-states $2$ and $1$. Each edge or path is labelled by the rates at which that specific family of states is produced; e.g.\ the state denoted by $\rho_{b,L}(2)$ (second from left) is produced with probability $\alpha\overline{\beta}$, while $I(R;L)=2$ with probability $\alpha$. $\overline{\alpha}\equiv1-\alpha$, and likewise for $\beta$ and $\gamma$}
    \label{fig:markovtree}
    \end{figure}
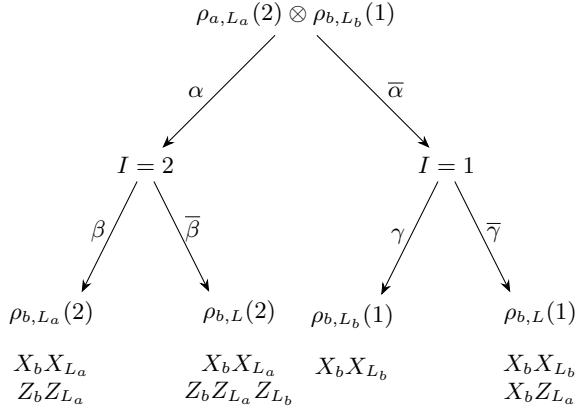

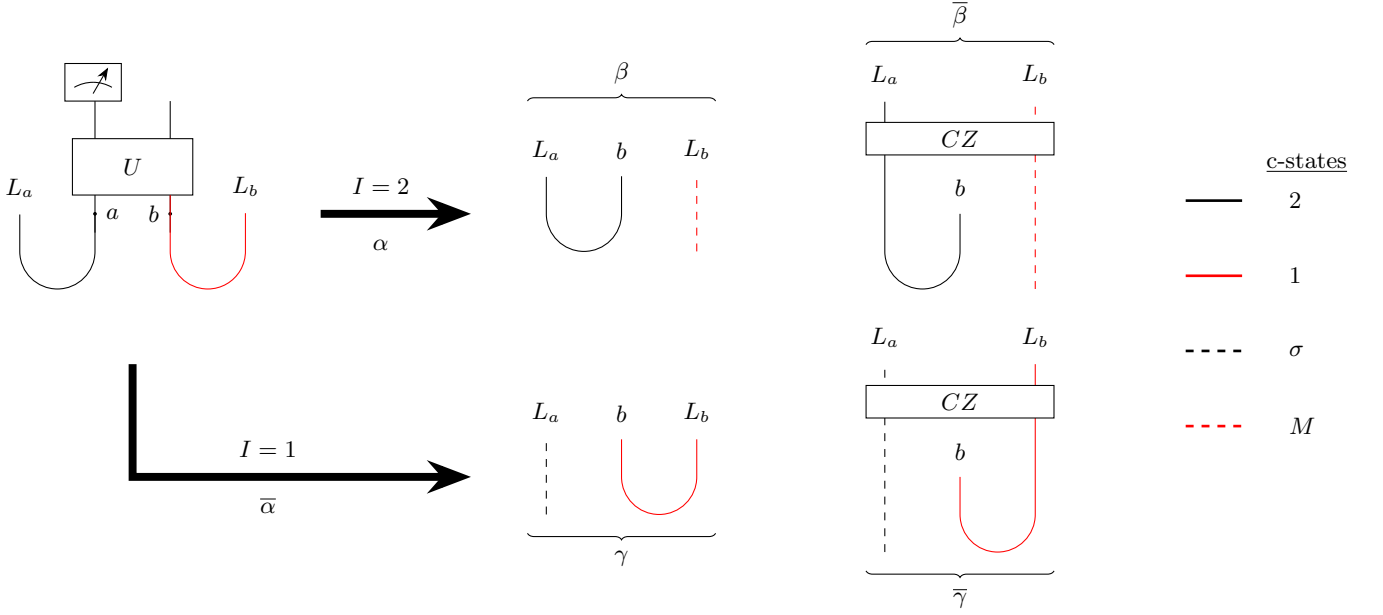
\begin{figure*}
        \begin{tikzpicture}
        \begin{scope}[xshift=-1cm]
            
    	\node[fill,circle,inner sep=0.5pt,label=right:$a$] at (0,0) {};
    	\node[fill,circle,inner sep=0.5pt,label=left:$b$] at (1,0) {};
    	
    	\repCirc[1];
    	
    	\draw (0,0) -- +(0,-0.5) arc[start angle = 360, end angle = 181,radius=0.5cm] -- +(0,0.5)node[label={[label distance=0cm]above:$L_a$}] {};
    	
    	\draw[red] (1,0.25) -- +(0,-0.75) arc[start angle = 180, end angle = 361,radius=0.5cm] -- +(0,0.5) node[label={[label distance=0.cm]above:\color{black} $L_b$}] {};
        \end{scope}

    \draw[line width = 3pt,->] (2,0) -- node[pos=0.4,align=center]{$I=2$\\ \\ $\alpha$} (4,0);
    \draw[line width = 3pt,->] (-0.5,-2) -- ++ (0,-1.5) -- node[pos=0.4,align=center]{$I=1$\\ \\ $\overline{\alpha}$} (4,-3.5);
	
	\begin{scope}[yshift=0.5cm,xshift=6cm]
		\draw (0,0) node[label=above:$b$]{} -- +(0,-0.5) arc[start angle = 360, end angle = 181,radius=0.5cm] -- +(0,0.5) node[label={[label distance=0cm]above:$L_a$}] {};
		
		\draw[red,dashed] (1,-1) -- +(0,1) node[black, label={[label distance=0cm]above:{\color{black}$L_b$}}] {};

        \draw[decorate,decoration=brace] (-1.25,1) -- node[label=above:$\beta$]{} (1.25,1);
	\end{scope}	
	
	\begin{scope}[xshift=10.5cm]
		\draw (0,0) node[label=above:$b$]{} -- +(0,-0.5) arc[start angle = 360, end angle = 181,radius=0.5cm] -- +(0,2) node[label={[label distance=0cm]above:$L_a$}] {};
		
		\draw[red,dashed] (1,-1) -- +(0,2.5) node[black, label={[label distance=0cm]above:{\color{black}$L_b$}}] {};
		
		\node[draw,fill=white,rectangle,inner xsep=1cm] at (0,1) {$CZ$};
        \draw[decorate,decoration=brace] (-1.25,2.25) -- node[label=above:$\overline{\beta}$]{} (1.25,2.25);
		
	\end{scope}
	
	\begin{scope}[yshift=-3cm,xshift=6cm]
		\draw[dashed] (-1,-1) -- +(0,1) node[label={[label distance=0cm]above:\color{black} $L_a$}]{};
		
		\draw[red] (0,0) node[label=above:{\color{black} $b$}]{} -- +(0,-0.5) arc[start angle = 180, end angle = 360,radius=0.5cm] -- +(0,0.5) node[label={[label distance=0cm]above:\color{black} $L_b$}] {};

        \draw[decorate,decoration={brace,mirror}] (-1.25,-1.25) -- node[label=below:$\gamma$]{} (1.25,-1.25);
		
	\end{scope}

	\begin{scope}[yshift=-3.5cm,xshift=10.5cm]
		\draw[dashed] (-1,-1) -- +(0,2.5) node[label={[label distance=0cm]above:\color{black} $L_a$}]{};
				
		\draw[red] (0,0) node[label=above:{\color{black} $b$}]{} -- +(0,-0.5) arc[start angle = 180, end angle = 360,radius=0.5cm] -- +(0,2) node[label={[label distance=0cm]above:\color{black} $L_b$}] {};
		
		\node[draw,fill=white,rectangle,inner xsep=1cm] at (0,1) {$CZ$};
        \draw[decorate,decoration={brace,mirror}] (-1.25,-1.25) -- node[label=below:$\overline{\gamma}$]{} (1.25,-1.25);
	\end{scope}

    \begin{scope}[xshift=13.5cm,line width=1pt,yshift=5pt,
			every node/.style={fill=none}]
			\path (0,0.5) -- +(0.75,0) node [right=0.2] {\underline{c-states}};
			\draw (0,0) -- (0.75,0) node [right=0.5] {$2$};
			\draw[red] (0,-1) -- (0.75,-1) node [color=black,right=0.5] {$1$};
			\draw[dashed] (0,-2) -- +(0.75,0) node[right=0.5]{$\sigma$};
			\draw[red,dashed] (0,-3) -- +(0.75,0) node[color=black,right=0.5] (M) {$M$};			
		\end{scope}
\end{tikzpicture}
\caption{Schematic representations of the circuits and example states considered in the definitions of $\alpha, \beta$ and $\gamma$, in \cref{fig:markovtree}. A pair of qubits $(a,L_a)$ in the c-state $I=2$ meet another pair of qubits $(b,L_b)$ -- which are in the $I=1$ state and therefore only classically correlated -- at a node. Following a unitary gate $U_{a,b}$, a measurement of $a$ and its subsequent tracing out, the result can be one of four families of states supported on $(b,L_b,L_a)$.}
\label{fig:markovtree_circ}
    \end{figure*}

    This Markov tree can alternatively be drawn in a circuit representation, as we have in \cref{fig:markovtree_circ}. While the visual similarity to tensor network diagrams is suggestive, we note that strictly speaking, \cref{fig:markovtree_circ} is not a collection of tensor networks. However, the translation between the two is straightforward.

	$\alpha, \beta$ and $\gamma$ are determined by the ensemble from which the unitary gates are drawn. The Markov trees in \cref{fig:markovtree,fig:markovtree_circ} further provide operational definitions of these parameters -- $\alpha$, for instance, labels the probability of the output state being maximally entangled with the leaves, when the inputs are the c-states $2$ and $1$. We will restrict our attention to ensembles that are invariant under the swapping of the two inputs. With these as starting points, we can now obtain $W((a,b)\to c)$ for arbitrary $a$ and $b$.

	\subsection{$a=M, b=1$}
	
	A maximally mixed state can be obtained by tracing the leaves out of $\rho_{a,L_a}(2)$, giving $\mathbb{1}_a \propto \Tr_{L_a}\qty[\rho_{a,L_a}(2)]$. Under a unitary $U$,
	\begin{equation}
		\begin{aligned}
			\mathbb{1}\otimes\rho_{b,L_b}(1)&\to\Tr_a\qty[\mathbb{P}_a U_{ab} \mathbb{1}\otimes\rho_{b,L_b}(1) U^\dagger_{ab} \mathbb{P}_a]\\
			&=\Tr_a\qty[\mathbb{P}_a U_{ab} \Tr_{L_a}\qty[\rho_{a,L_a}(2)]\otimes\rho_{b,L_b}(1)  U^\dagger_{ab} \mathbb{P}_a]\\
			&=\Tr_{L_a}\Tr_a\qty[\mathbb{P}_a U_{ab} \rho_{a,L_a}(2)\otimes\rho_{b,L_b}(1)  U^\dagger_{ab} \mathbb{P}_a],
		\end{aligned}
	\end{equation}
	where the last step follows from the fact that the circuit does not directly act on the leaves, allowing the partial trace over the leaves to commute with the unitary gate and subsequent measurement. However, the quantity in the square brackets in the last line
	\[\Tr_a\qty[\mathbb{P}_a U_{ab} \rho_{a,L_a}(2)\otimes\rho_{b,L_b}(1)  U^\dagger_{ab} \mathbb{P}_a]\]
	is \textit{exactly} the output quantum state when the inputs are $\rho(2)$ and $\rho(1)$. Therefore, the outputs when $a=M$ can be obtained by tracing $L_a$ out in the outputs corresponding to the case where the input was $a=2$. Represented graphically,
	
	\begin{tikzpicture}[level 1/.style = {level distance=2.5cm, sibling distance=4cm}]
		\node [align=center] {$\begin{aligned}
				&\qquad\mathbb{1}\otimes\rho_{b,L_b}(1)\\
				=&\Tr_{L_a}\qty[\rho_{a,L_a}(2)]\otimes\rho_{b,L_b}(1)
			\end{aligned}$}
		child {node {$\begin{aligned}
						\Tr_{L_a}&\qty[\rho_{a,L}(2)]\\
						&\propto \mathbb{1}
				\end{aligned}$}
			edge from parent [->] node [pos=0.7,label=above:$\alpha$] {}
			}
		child { node {$\begin{aligned}
					\Tr_{L_a}&\qty[\rho_{b,L}(1)]\\
					&= \rho_{b,L_b}(1)
			\end{aligned}$
		}edge from parent [->] node [pos=0.7,label=above:$\overline{\alpha}$] {}
		};
	\end{tikzpicture}
	
	Thus,
	
	\begin{equation}
		\begin{aligned}
			W((M,1)\to M) &= \alpha\\
			W((M,1)\to 1) &= \overline{\alpha}
		\end{aligned}.
	\end{equation}
	
	With the basic strategy clear, we will use this graphical language to explain the emergence of $W((a,b)\to c)$ from $W((2,1)\to \dots)$.
	
	\subsection{$a=2, b=\sigma$}
	
	Instead of using a partial trace, we projectively measure $L_b$ in $\rho_{a,L_b}(1)$ (in a basis that commutes with $\rho(1)$) to obtain the product state $\rho_b(\sigma)$ on $b$. Explicitly, we can assume that $\rho_{b,L_b}(1) \propto \mathbb{1} + Z_bZ_{L_b}$, so that $\rho_b(\sigma) = \mathbb{P}_{L_b}\rho_{b,L_b}(1)\mathbb{P}_{L_b}$, where $\mathbb{P}_{L_b}$ enacts the projection onto a $Z$ basis state on $L_b$. Again, since the gates and measurements on the node do not explicitly involve $L_b$, the projection operator $\mathbb{P}_{L_b}$ commutes with them, as $\Tr_{L_a}$ did. Moreover, since the only stabilizer supported on $L_b$ is $Z_{L_b}$, the stabilizers of the output state can only either include $Z_{L_b}$ or $\mathbb{1}_{L_b}$ on the leaf $L_b$. This can be seen by explicitly considering the evolution of the stabilizers of $\rho(2)\otimes\rho(1)$ under any $U_{ab}$ and a measurement on $a$ -- $U_{ab}$ by definition does not change the action of any stabilizer on $L_b$. Moreover, under measurements, some stabilizers have to be multiplied with each other or discarded entirely; this ensures that the action of any stabilizer on $L_b$ is $Z^s_{L_b},\text{ }s=0,1$. Proceeding as before,\\

	\begin{tikzpicture}[level 1/.style = {level distance=2.5cm, sibling distance=4cm}]

		\node [align=center] {$\begin{aligned}
				&\qquad\rho_{a,L_a}(2)\otimes\rho_{b,L_b}(\sigma)\\
				=&\text{ }\rho_{a,L_a}(2)\otimes\mathbb{P}_{L_b}\rho_{b,L_b}(1)\mathbb{P}_{L_b}
			\end{aligned}$}
		child {node {$\begin{aligned}
					&\mathbb{P}_{L_b}\rho_{b,L}(2)\mathbb{P}_{L_b}\\
					&\propto \rho_{b,L}(2)
				\end{aligned}$}
			edge from parent [->] node [pos=0.7,label=above:$\alpha$] {}
		}
		child { node {$\begin{aligned}
					&\mathbb{P}_{L_b}\rho_{b,L}(1)\mathbb{P}_{L_b}\\
					&= \rho_{b}(\sigma)
				\end{aligned}$
			}edge from parent [->] node [pos=0.7,label=above:$\overline{\alpha}$] {}
		};
	\end{tikzpicture}\\

	That $\mathbb{P}_{L_b}\rho_{b,L}(1)\mathbb{P}_{L_b} = \rho_b(\sigma)$ follows from the stabilizer tableau of $\rho_{b,L_b}(1)$ having $P_b Z_{L_b}$ alone or $\qty{P_b Z_{L_b}, P_b P^{'}_{L_a}}$, where $P, P'$ are arbitrary Pauli operators. Upon measuring $Z_{L_b}$, the observable commutes with both stabilizers and is therefore added to the generating stabilizer set. Simplifying this modified set results in a new set consisting of the operators $\qty{P_b, Z_{L_b}}$, showing that the resulting state on $b$ is indeed a product state.
	
	As we found previously, 
	\begin{equation}
		\begin{aligned}
			W((2,\sigma)\to 2) &= \alpha,\\
			W((2,\sigma)\to \sigma) &= \overline{\alpha}.
		\end{aligned}
	\end{equation}
	
	\subsection{$a=M, b=\sigma$}
	
	We demonstrated in \cref{sec:mipt} how $W((M,\sigma))$ can be obtained from the transition matrix for $a=2, b=\sigma$, since the product state $\sigma$ is agnostic to the exact qubit to which qubit $a$ is entangled. We restate for completeness that 
	
	\begin{equation}
		\begin{aligned}
			W((M,\sigma)\to M) &= \alpha,\\
			W((M,\sigma)\to \sigma) &= \overline{\alpha}.
		\end{aligned}
	\end{equation}
	
	\subsection{$a=1, b=\sigma$}
	
	In a slight departure from previous derivations, we use the case of $a=2, b=\sigma$ as a starting point for this case. Recall from \cref{eq:eg_I1} in the main text that one purification of an $I=1$ state on $a$ is a GHZ state between $a$, $L_a$ and an environment, which we label as $e$. We denote by $\rho_{a,L_a,e}$ the pure state from which $\rho_{a,L_a}(1) \equiv \Tr_e\qty[\rho_{a,L_a,e}]$ can be obtained by partial tracing. The stabilizer set of $\rho_{a,L_a,e}$ is $\qty{X_aX_{L_a}X_e, Z_aZ_{L_a}, Z_{L_a}Z_e}$. $\rho_{a,L_a,e}$ can be disentangled from $e$ by the application of a $CNOT$ (hereafter denoted $CX$) gate on $\qty{L_a,e}$, resulting in
	
	\begin{equation}
		\begin{aligned}
			CX \rho_{a,L_a,e} CX &= \rho_{a,L_a}(2) \otimes \rho_e(\sigma)\\
			\implies \rho_{a,L_a}(1) &= \Tr_e\qty[\rho_{a,L_a,e}]\\
			&= \Tr_e\qty[CX \rho_{a,L_a}(2)\otimes \rho_e(\sigma)CX]
		\end{aligned}
	\end{equation}
	
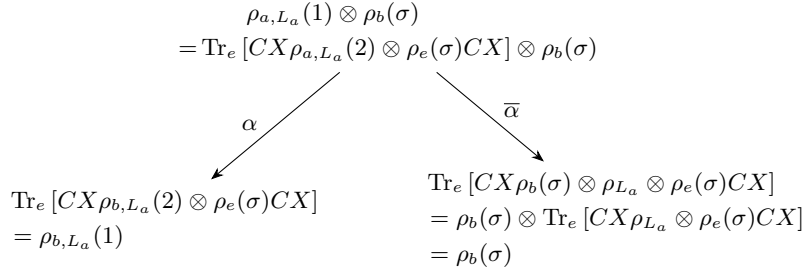
\begin{figure*}
        \begin{tikzpicture}[level 1/.style = {level distance=2.5cm, sibling distance=6cm}]
		
		\node [align=center] {$\begin{aligned}
				&\qquad\rho_{a,L_a}(1)\otimes\rho_{b}(\sigma)\\
				=&\Tr_e\qty[CX \rho_{a,L_a}(2)\otimes \rho_e(\sigma)CX]\otimes\rho_b(\sigma)
			\end{aligned}$}
		child {node {$\begin{aligned}
					&\Tr_e\qty[CX \rho_{b,L_a}(2)\otimes\rho_e(\sigma) CX]\\
					&= \rho_{b,L_a}(1)
				\end{aligned}$}
			edge from parent [->] node [pos=0.7,label=above:$\alpha$] {}
		}
		child { node {$\begin{aligned}
					&\Tr_e\qty[CX \rho_{b}(\sigma)\otimes\rho_{L_a}\otimes\rho_e(\sigma) CX]\\
					&= \rho_{b}(\sigma)\otimes\Tr_e\qty[CX\rho_{L_a}\otimes\rho_e(\sigma) CX]\\
					&= \rho_b(\sigma)
				\end{aligned}$
			}edge from parent [->] node [pos=0.7,label=above:$\overline{\alpha}$] {}
		};
	\end{tikzpicture}
        \caption{Markov tree for the process where the inputs are a product state $\sigma$ and a mixed state that is only classically correlated with its leaves, represented by the c-state $1$.}
        \label{fig:a1bs}
        \end{figure*}

	The value of $\sigma$ in $\rho_e(\sigma)$  is explicitly specified to be $Z$, since the $e$ qubit is an eigenstate of $Z$. However, this has no impact on the generality of our results, since any other purification could also have been chosen. In that case, the disentangling unitary would not be $CX$, but a different two-qubit gate, controlled nonetheless on $L_a$. Since $\Tr_e$ and $CX_{L_a,e}$ both commute with the unitaries on the node (but not each other), we can apply them after the unitaries and measurements on $a,b$ have taken place. The Markov tree is presented in \cref{fig:a1bs}.

	We have written 
    \[\rho_{b}(\sigma)\otimes\Tr_e\qty[CX_{L_a,e}\rho_{L_a}\otimes\rho_e(\sigma) CX_{L_a,e}] = \rho_b(\sigma),\] since the output qubit is completely disentangled from the leaves and the environment $e$. The transition probabilities are
    \begin{equation}
		\begin{aligned}
			W((1,\sigma)\to 1) &= \alpha,\\
			W((1,\sigma)\to \sigma) &=\overline{\alpha}.
		\end{aligned}
	\end{equation}
	
	\subsection{$a=2, b=M$}
	
\begin{figure*}
    \begin{tikzpicture}[
		level 1/.style = {level distance=2cm, sibling distance=6cm},
		level 2/.style = {sibling distance=3cm, level distance=2.5cm}]
		\node (R) {$\begin{aligned}
				\rho_{a,L_a}(2)&\otimes \mathbb{1}_b\\
				\propto\rho_{a,L_a}(2)&\otimes\Tr_{L_b}\qty[\rho_{b,L_b}(1)]
			\end{aligned}$}
		child { node (I2) {$\Tr_{L_b}\qty[\rho_{b,L}(2)]$} 
			child { node[align=center] (I2a) {$\begin{aligned}
						\Tr_{L_b}&\qty[\rho_{b,L_a}(2)]\\
						=&\rho_{b,L_a}(2)
					\end{aligned}$}
				edge from parent [->] node [pos=0.7,label=above:$\beta$] {}}
			child { node[align=center] (I2b) {$\begin{aligned}
						\Tr_{L_b}&\qty[\rho_{b,L}(2)]\\
						=&\rho_{b,L_a}(1)
					\end{aligned}$}
				edge from parent [->] node [pos=0.7,label=above:$\overline{\beta}$] {}}
			edge from parent [->] node [pos=0.7,label=above:$\alpha$] {}
		}	
		child {node (I1) {$\Tr_{L_b}\qty[\rho_{b,L}(1)]$}
			child { node[align=center] (I1a) {$\begin{aligned}
						\Tr_{L_b}&\qty[\rho_{b,L_b}(1)]\\
						=&\mathbb{1}_b
					\end{aligned}$
				}
				edge from parent [->] node [pos=0.7,label=above:$\gamma$] {}}
			child { node[align=center] (I1b) {$\begin{aligned}
						\Tr_{L_b}&\qty[\rho_{b,L}(1)]\\
						=&\rho_{b,L_a}(1)
					\end{aligned}$}
				edge from parent [->] node [pos=0.7,label=above:$\overline{\gamma}$] {}}
			edge from parent [->] node [pos=0.7,label=above:$\overline{\alpha}$] {}
		};
	\end{tikzpicture}
    \caption{Markov tree for the process where the inputs are a maximally mixed state $M$ and a state that forms a Bell pair with its leaves, represented by the c-state $2$. Note the inequivalent treatments of the two different types of $\rho(2)$ and $\rho(1)$ states, which bring the role of $\beta$ and $\gamma$ to the fore.}
        \label{fig:a2bm}
    \end{figure*}
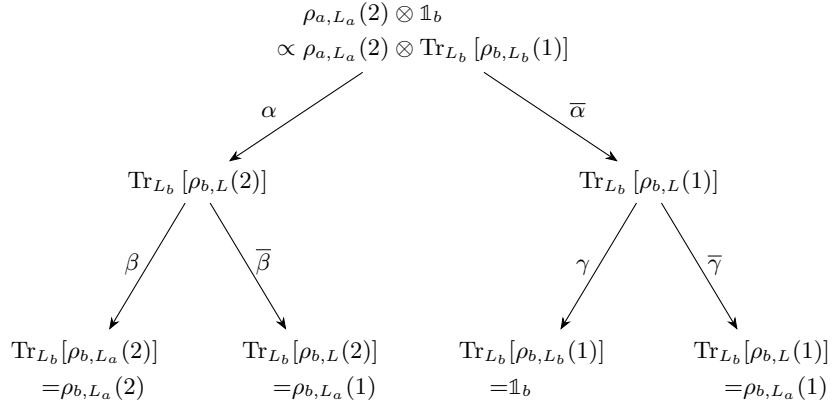

	The last case we consider is when one state is maximally entangled to the leaves, while the other is maximally entangled to the environment (or equivalently, maximally mixed, since the state of the environment is inaccessible under generic conditions). Our starting point is now the case of $a=2, b=1$, noting that $\Tr_{L_b}\qty[\rho_{b,L_b}(1)] = \mathbb{1}_b$. This is the only case where the parameters $\beta$ and $\gamma$ feature explicitly in $W$. The Markov tree is shown in \cref{fig:a2bm}.

	The cases that result in an output of an $I=1$ state can be explained as follows. The generators of the stabilizers of both $\rho_{a,L}(2)$ and $\rho_{b,L}(1)$\footnote{Note that the leaves are supported on $L=L_a\cup L_b$ and not $L_{a,b}$ alone} can be cast in a form where only one generator has support on $L_b$, with the other supported on $b$ and $L_a$ alone. Therefore, tracing $L_b$ out removes that stabilizer which has a support on $L_b$. Since what remains (in either case) is a Clifford state on two qubits, but with only one nontrivial (i.e.\ non-identity) stabilizer, this state must correspond to an $I=1$ state. Moreover, the state $\rho_{b,L_b}(1)$ has only one nontrivial stabilizer of the form $P_bP^{'}_{L_b}$, so tracing $L_b$ out naturally results in a maximally mixed state $\mathbb{1}_b$.
	
	This gives us that
	
	\begin{equation}
		\begin{aligned}
			W((2,M)\to2) = \alpha\beta,\\
			W((2,M)\to1) = \alpha\overline{\beta} + \overline{\alpha}\overline{\gamma},\\
			W((2,M)\to M) = \overline{\alpha}\gamma.
		\end{aligned}
	\end{equation}
	
	The transition probabilities are all summarized in \cref{tab:WMat}.
	
	\begin{table}
		\centering
		\begin{tblr}{colspec={cccccc},hlines,vlines,
				hline{1-4,14} = {1pt},vline{2-6} = {1-3}{1pt},
				vline{3} = {1pt},vline{1,7}={1pt},
			row{1,2}={font=\bfseries}}
			\SetCell[c=2]{c}&&\SetCell[c=4]{c} $W((a,b)\to c)$&&&\\
			\SetCell[r=2]{c}a&\SetCell[r=2]{c}b&\SetCell[c=4]{c} c&&&\\
			&&2&1&$\sigma$&$M$\\
			2&2&1&0&0&0\\
			2&1&$\alpha$&$\overline{\alpha}$&0&0\\
			2&$\sigma$&$\alpha$&0&$\overline{\alpha}$&0\\
			2&$M$&$\alpha\beta$&$\alpha\overline{\beta} + \overline{\alpha}\overline{\gamma}$&0&$\overline{\alpha}\gamma$\\
			1&1&0&1&0&0\\
			1&$\sigma$&0&$\alpha$&$\overline{\alpha}$&0\\
			1&$M$&0&$\overline{\alpha}$&0&$\alpha$\\
			$\sigma$&$\sigma$&0&0&1&0\\
			$\sigma$&$M$&0&0&$\overline{\alpha}$&$\alpha$\\
			$M$&$M$&0&0&0&1\\
		\end{tblr}
		\caption{The transition probabilities $W((a,b)\to c)$ for all unique pairs $(a,b)$. Note that all the ensembles considered in this work are symmetric under the interchanging of their inputs, i.e.\ $W((a,b)\to c) = W((b,a)\to c)$. The two columns on the left contain the inputs, while the remaining columns contain the transition probabilities to the respective states $c$.}
		\label{tab:WMat}
	\end{table}

\section{Parametrization of Deterministic Gates}
\label{app:a0a1gates}

In this appendix, we demonstrate how the values of $\alpha, \beta$ and $\gamma$ are obtained for the two gates presented in \cref{fig:tuningA} in the main text. These parameters can be determined solely by observing the action of these gates on qubits in specific c-states. Unlike the Clifford ensembles considered earlier, these gates are \textit{not} invariant under single qubit rotations, so we need to specify the initial state (taken to be Bell pairs with the stabilizers $X_aX_L$ and $Z_aZ_L$).

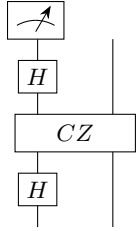
\begin{figure}
	\centering
	\begin{tikzpicture}
		\foreach \x in {0,1cm}
		{
			\draw[xshift=\x] (0,-0.5) -- (0,2);
		}
		
		\draw[fill=white] (-0.3,0.5) rectangle node {$CZ$} (1.3,1);
		\node[rectangle,draw,fill=white] at (0,1.5) {$H$};
		\node[draw,fill=white] at (0,0) {$H$};
		\meas[(-0.4,2)]
	\end{tikzpicture}
	\caption{A deterministic gate with $\alpha=0$.}
	\label{fig:a0_gate}
\end{figure}

First, we consider the gate presented in \cref{fig:a0_gate}. Letting the c-state on $a$ be 1, and on $b$ be $M$, we trace their evolution. Recall that one definition of $\alpha$ is $W((1,M)\to M)\equiv \alpha$. In what follows, $P_L$ is any Pauli operator on $L$, and we omit non-identity stabilizers.

\begin{center}
	\begin{tikzpicture}[node distance=1cm]
		\node[align=center] (n1) at (0,0) { $Z_aP_{L_a}$};
		\node[right=of n1, align=center] (n2) { $X_aP_{L_a}$};
		\node[right=of n2, align=center] (n3) { $Z_bX_aP_{L_a}$};
		\node[below=of n3, align=center,inner sep=6pt] (n4) {$Z_bX_aP_{L_a}$\\$X_a$};
		\node[left=of n4, align=center] (n5) {$Z_bP_{L_a}$};
		
		\draw[->] (n1.east) -- node[pos=0.5,label={[label distance=-0.2cm]above:$H_a$}]{} (n2.west);
		\draw[->] (n2.east) -- node[pos=0.5,label={[label distance=-0.2cm]above:$CZ$}]{} (n3.west);
		\draw[->] (n3.south) -- node[pos=0.5,align=center,label=right:Measure $X_a$]{} (n4.north);
		\draw[->] (n4.west) -- node[pos=0.5,label={[label distance=-0.2cm]above:$\tr_a$}]{} (n5.east);
	\end{tikzpicture}
\end{center}

The output is a mixed state, classically correlated (c-state 1) with the leaves $L_a$, so $\alpha=0$ for this gate.

\begin{figure}
	\centering
	\begin{tikzpicture}
		\foreach \x in {0,1cm}
		{
			\draw[xshift=\x] (0,-0.5) -- (0,2);
		}
		
		\draw[fill=white] (-0.3,0.5) rectangle node {$CZ$} (1.3,1);
		\node[rectangle,draw,fill=white] at (0,1.5) {$H$};
		\node[draw,fill=white] at (1,0) {$H$};
		\meas[(-0.4,2)]
	\end{tikzpicture}
	\caption{A deterministic gate with $\alpha=1$.}
	\label{fig:a1_gate}
\end{figure}
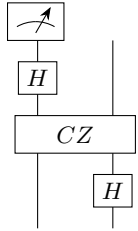

For the other gate (\cref{fig:a1_gate}), its action on the same c-states is

\begin{tikzpicture}
	\node[align=center] (n1) at (0,0) { $Z_aP_{L_a}$};
	\node[right=of n1, align=center] (n2) { $Z_aP_{L_a}$};
	\node[right=of n2, align=center] (n3) { $Z_aP_{L_a}$};
	\node[below=of n3, align=center] (n4) {$X_a$};
	\node[left=of n4, align=center] (n5) {$\mathbb{1}_b$};
	
	\draw[->] (n1.east) -- node[pos=0.5,label={[label distance=-0.2cm]above:$H_b$}]{} (n2.west);
	\draw[->] (n2.east) -- node[pos=0.5,label={[label distance=-0.2cm]above:$CZ$}]{} (n3.west);
	\draw[->] (n3.south) -- node[pos=0.5,align=center,label=right:Measure $X_a$]{} (n4.north);
	\draw[->] (n4.west) -- node[pos=0.5,label={[label distance=-0.2cm]above:$\tr_a$}]{} (n5.east);.
\end{tikzpicture}

The output here is a maximally mixed state, so $\alpha=1$ for this gate. Following this exercise for input states $2$ and $1$ instead will show that $\beta=\gamma=0$ for both these gates. Moreover, through explicit stabilizer calculations, the outputs are known to remain in a manifold of states spanned by 
\begin{equation}
	\begin{aligned}
		\rho_{a,L_a}(2) &\propto U_{L_a} \qty(\mathbb{1} + X_a X_{L_a})\qty(\mathbb{1} + Z_a Z_{L_a})U^\dagger_{L_a}\\
		\rho_{a,L_a}(1) &\propto U_{L_a} \qty(\mathbb{1} + Z_a Z_{L_a})U^\dagger_{L_a}\\
		\rho_{a,L_a}(M) &\propto \mathbb{1}_a.
	\end{aligned}
\end{equation}

$U_{L_a}$ is an arbitrary single-qubit Clifford gate that acts on the leaves alone, and is hence inconsequential to our results.

\bibliography{reference}
\end{document}